\documentclass[11pt,oneside]{article}
\usepackage{setspace,graphicx,amssymb,amsmath,amsfonts,amscd,amsthm,multirow,mathdots,caption,array}
\usepackage{pdflscape,empheq}
\usepackage{tabularx,cite,mathrsfs,color}
\usepackage{stmaryrd}
\usepackage{rotating}
\usepackage{authblk}
\usepackage{hyperref}
\usepackage{float}
\usepackage{ctable}
\usepackage{geometry}
\usepackage{geometry}

\usepackage[font=small,labelfont=bf]{caption}

\newcommand{\bea}{\begin{eqnarray}}
\newcommand{\eea}{\end{eqnarray}}
\newcommand{\bee}{\begin{eqnarray*}}
\newcommand{\eee}{\end{eqnarray*}}
\newcommand{\al}{\begin{align*}}
\newcommand{\eal}{\end{align*}}
\newcommand{\be}{\begin{equation}}
\newcommand{\ee}{\end{equation}}
\newcommand{\eq}[1]{(\ref{#1})}
\newcommand{\bem}{\begin{pmatrix}}
\newcommand{\eem}{\end{pmatrix}}

\def\a{\alpha}

\def\d{\delta}

\def\m{\mu}
\def\n{\nu}

\def\o{\omega}

\def\pa{\partial}

\def\s{\sigma}

\def\til{\tilde}

\newcolumntype{R}{ >{$}r <{$}}
\newcolumntype{C}{ >{$}c <{$}}
\newcolumntype{L}{ >{$}l <{$}}
\newcolumntype{F}{>{\centering\arraybackslash}m{1.5cm}}

\newcommand{\comment}[1]{}

\newcommand{\RR}{{\mathbb R}}%Reals
\newcommand{\ZZ}{{\mathbb Z}}%Integers
\theoremstyle{definition}

\theoremstyle{remark}

\numberwithin{equation}{section}
\title{
\vspace{-35pt}
    \textsc{\Large{
Flux vacua: A voluminous recount
            }  }
}

\author[1,2]{\small{Miranda C. N. Cheng}\thanks{On leave from CNRS, France.}}
\author[3]{\small{Gregory W. Moore}}
\author[4]{\small{Natalie M. Paquette}}

\date{}

\affil[1]{Korteweg-de Vries Institute for Mathematics, Amsterdam, the Netherlands}
\affil[2]{Institute of Physics, University of Amsterdam, Amsterdam, the Netherlands}
\affil[3]{NHETC and Department of Physics and Astronomy, Rutgers University, 126 Frelinghuysen Rd., Piscataway NJ 08855, USA }
\affil[4]{Walter Burke Institute for Theoretical Physics, California Institute of Technology, Pasadena, CA
91125, USA}

\begin{document}

 \pagenumbering{gobble}

\setstretch{1.4}

   \maketitle
   \abstract
   {
In this note we apply mathematical results for the volume of certain symmetric spaces to the problem of  counting flux vacua in simple IIB Calabi--Yau compactifications. In particular we obtain estimates for the number of flux vacua \emph{including the geometric factor related to the Calabi-Yau moduli space}, in the large flux limit, for the FHSV model and some closely related models. We see that these geometric factors give rise to contributions to the counting formula that are typically not of order one and might potentially affect the counting qualitatively in some cases. We also note, for simple families of Calabi-Yau moduli spaces, an interesting dependence of the moduli space volumes on the dimension of the flux space, which in turn is governed by the Betti numbers of the Calabi-Yaus.

\bigskip

\bigskip

\hfill \emph{CALT-TH-2019-033}
}

\newpage
\pagenumbering{arabic}
\tableofcontents

%\setstretch{1.4}

%\title{
%\vspace{-35pt}
   % \textsc{\large{
     %Flux vacua: A voluminous recount
        %    }  }
%}

%   \maketitle
%\abstract{
%In this note we apply mathematical results for the volume of certain symmetric spaces to the problem of  counting flux vacua in simple IIB Calabi--Yau compactifications. In particular we obtain estimates for the number of flux vacua \emph{including the geometric factor related to the Calabi-Yau moduli space}, in the large flux limit, for the FHSV model and some closely related models. We see that these volume factors give rise to contributions to the counting formula that are typically not of order one and might affect the counting qualitatively in some cases. We also note an interesting correlation between the dimension of the flux space and the size of the moduli space in this specific family of Calabi--Yau compactifications.
%}

%\tableofcontents

\section{Introduction}\label{intro}

The purview of this note is to re-count flux vacua in certain simple string compactifications. The string theory landscape, the huge number of vacua arising from string theory, is  among the most influential and controversial concepts in string theory, or even high-energy physics, in the past decade. The purported existence of such a landscape suggests the possibility that the fine-tuning of fundamental constants like the cosmological constant or the Higgs boson mass may be explained not by a physical mechanism but rather by a statistical argument, which is appropriate under the assumption that many ``universes" are equally consistent from the point of view of the fundamental laws of physics despite looking nothing like ours.

Given the important consequences, in this note we revisit the counting formula and in particular we focus on a factor,
which we will call the \emph{geometric factor}, that is often taken to be of order one in the literature.
The main question that motivates this note, and which we answer in specific cases, is whether the
geometric factor really is of order one, or not. The answer will be: ``Sometimes yes, sometimes no.''
To be precise, the geometric factor is the expression  $\pi^{-m/2}\int \det(\mathcal{R} + \omega\cdot{\bf 1})$, in
equation \eqref{eq:IndexVac}.

This question is a reasonable one because existing computations of volumes of string theory moduli spaces
with respect to the Zamolodchikov metric have yielded numbers that, depending on the context, have been prodigious or miniscule.
\footnote{We have in mind the computations in \cite{Moore}, which provided some motivation for the present work.
The work of \cite{Moore}, related to a question posed in \cite{BCKMP}, used these volumes to estimate the likelihood that certain CFTs have a weakly-curved AdS gravity duals. Moduli spaces associated with superconformal field
theories built from products of $\textrm{Hilb}^n(K3)$ and $\textrm{Hilb}^n(T^4)$ were considered and the corresponding
Zamolodchikov volumes were found to be extremely small in examples relevant to string compactification.}
Of course, the answer depends sensitively on the choice of metric on the space in question. In particular, since the dimensionality of Calabi-Yau complex structure moduli spaces can be enormous (see \cite{Taylor:2015xtz} for a particularly striking recent example), obtaining the correct normalisation of the metric is essential: rescaling the metric by a numerical factor $\lambda$ will scale the volume by a factor $\lambda^{\textrm{dim}_{\mathbb{R}}(\mathcal{M})/2}$.

It is very hard to analyze the geometric factor for Calabi-Yau threefolds of generic holonomy. The global form of the Calabi-Yau moduli spaces is not known in general, and the curvature of the space, present in the geometric factor, is not uniform over moduli space; for instance, the curvature is known to diverge (albeit remains integrable!) in the vicinity of the conifold point; see \cite{Denef:2004ze, GKT, ConlonQuevedo, EguchiTachikawa} for an exploration of the geometric factor in certain IIB vacua in the vicinity of the conifold point, as well as in the neighborhoods of other tractable regimes in moduli space. Yet for specific, and indeed quite special Calabi--Yau manifolds, we are able to compute this factor, essentially given by the volume of the moduli space, exactly, by applying certain recent mathematical results. For these examples, we see that the answer to the question can be either yes or no. We report on the effect of the exact geometric factor for the case of the FHSV Calabi-Yau manifold in Table \ref{tbl:FHSV}, whereupon the geometric term contributes a factor of $10^{-8}$ (significant but still subleading in the limit of large flux). On the other hand, one can more properly consider orientifolds of this model. The simplest choice of orientifold action dramatically reduces the dimension of the moduli space, whereupon its volume corrects the volume estimate by a paltry $10^{-2}$.

In Section \ref{sec:estimates} we make a curious observation: If we consider certain families of moduli spaces of
increasing dimension then the volume is a steeply decreasing and then increasing function of the dimension and
the minimal value can be extremely small. Moreover, the minimum appears at the dimensions most relevant for
string compactification. (A similar phenomenon occurs with $(4,4)$ sigma models \cite{Moore}.)
While this might well be an artifact of the examples we have considered it might also be more general.
If so, it could have important consequences for the main claim of \cite{Taylor:2015xtz}.

We begin by recalling the flux vacua counting formula \cite{AshokDouglas}, which builds on the seminal work of \cite{BoussoPolchinski}. Our exposition will closely follow  that of \cite{Denef}.
Consider a region $S$ in a space with real coordinates $x^i$, $i=1,\dots, m$, equipped with a K\"ahler structure. For our application this will be (a region of) the complex structure moduli space of the F-theory fourfold.
Let $P_{Ii}$, $I=1,\dots,b$ be a set of real vector fields \footnote{Per \cite{Denef}, the derivation is presented assuming $x^i, P_{Ii}$ are real, but the argument goes through with minor modifications when $x^i, P_{Ii}$ are complex. In the F-theory context, they are to be identified, respectively, with the coordinates on complex structure moduli space and derivatives of the period vector; see \cite{Denef} for the precise identifications.} and let $A_{IJ}$ be given by a non-degenerate, symmetric bilinear form. In F-theory, $A_{IJ} = -Q_{IJ}$, where the latter is the intersection product on the integral homology lattice of the fourfold. For a given $L_{\rm max}$, we would like to count the number of pairs $(\underline N, x_\ast )$ where
\be\label{eq:fluxes} \underline N = (N_1,\dots, N_b), ~ N^I \in \ZZ ~~{\rm satisfying}~~
\tfrac{1}{2} N^I N^J A_{IJ} \leq L_{\rm max} = {R_{\rm max}^2\over 2}
\ee
and $x_\ast  \in S$ such that
$U_{\underline N;i}:=\sum_I N^I P_{Ii} =0$ for all $i$.
It is easy to see that such a number is given by
\be
N_{\rm zeros} =\sum_N \int_S d^mx\,\left(\prod_i \d(U_{\underline N;i})\right) \, \lvert{\rm det}(\pa_j U_{\underline N;k})\rvert ,
\ee
where the sum is taken over fluxes satisfying (\ref{eq:fluxes}). Assuming that there is no large cancellation and the absolute value $\lvert{\rm det}(\pa_j U_{\underline N;i})\rvert$ can be replaced by ${\rm det}(\pa_j U_{\underline N;i})$,
in the limit where the discreteness of $\underline N$ can be ignored the above quantity is approximated by the index
\be
I_{\rm zeros} = \int d^b N \int_S d^mx\,\left(\prod_i \d(U_{\underline N;i})\right) \, {\rm det}(\pa_j U_{\underline N;k}),
\ee
which, in the present context of flux vacua, can be shown to be the same as
\be\label{eq:fluxcount}
I_{\rm zeros} = \frac{1}{\sqrt{{\rm det}A}} \,
{\rm vol}_{R_{\rm max}}({\mathbb B}^{b})\,
 \int_S \frac{{\rm det} ({\cal R}+\omega\cdot {\bf 1})}{\pi^{m/2}},
\ee
where ${\cal R}$ is  the Ricci curvature of the holomorphic tangent bundle and $\o$ is the K\"ahler form of the
Weil-Peterson metric on $S$, and
$${\rm vol}_{R_{\rm max} }({\mathbb B}^{b}) ={ (2\pi L_{\rm max})^{b/2}\over (b/2)!}$$
 is the volume of the $b$-dimensional ball of radius $R_{\rm max} =\sqrt{2L_{\rm max}} $.

Under the above assumptions, and setting aside the question of K\"ahler  moduli stabilisation (by assuming that the moduli are stabilised by quantum effects), the number of vacua in type IIB flux compactification is given by
\be\label{eq:IndexVac}
I_{\rm vac} ={\rm vol}_{R_{\rm max}}({\mathbb B}^{b})
 \int_S \frac{{\rm det} ({\cal R}+\omega\cdot {\bf 1})}{\pi^{m/2}},
\ee
where we have used the fact that the bilinear form $A$ is given by the intersection form and has determinant 1. In terms of the F-theory data and in particular the fourfold $Y$, we have in the above formula $b={\rm dim}_\RR H$, where $H\subset H^4(Y,\RR)$ is space of  all $G \in  H^4(Y,\RR)$ satisfying $\int_Y G\wedge D\wedge D'=0$ for all $D, D'\in H^{1,1}(Y,\RR)$. It is not hard to see that $b$, being the dimension of the subspace of $H^4(Y)$ orthogonal to intersections of divisors, is equivalent to the dimension of the subspace of fluxes with exactly one leg in the elliptic fibre. The maximal number of fluxes is given by the tadpole cancellation condition
\footnote{Notice that although $A_{IJ}$ is a form of indefinite signature, the restriction to the set  of $N$ that
admit a supersymmetric vacuum is positive definite, and therefore the tadpole constraint
does bound the region of allowed fluxes $N$.  Furthermore, though one can reduce the upper bound on flux slightly by adding anti-D3 branes, one cannot add an arbitrary number of these: a sufficient number of anti-D3 branes in a flux background will decay to a configuration that contains only flux and D3-branes \cite{KPV}.}
\be\label{eq:tadpole}
\frac{1}{2}A_{IJ}N^I N^J + N_{D3} = \frac{\chi(Y)}{24} \Rightarrow
L_{\rm max} = \frac{\chi(Y)}{24}.
\ee

The vacua counting formula we will use in this note is obtained from the above by making extra assumptions, as in \cite{AshokDouglas}.
Namely, we consider the number of bulk flux vacua in the weakly coupled type IIB limit and ignore the D7 degrees of freedom.
Let $X$ be the Calabi--Yau threefold in the type IIB orientifold compactification and $n=h_{2,1}^-(X)$ to be the dimension of the subspace of $H^{2,1}(X,\mathbb Z)$ that is anti-invariant under the orientifold action.
In this limit the four-fold can be taken to be $Y=(T^2\times X)/{\mathbb Z}_2$ and  we have $b/2=2n+2$, corresponding to the $(n+1)$ Ramond--Ramond and $(n+1)$ NS--NS fluxes one can turn on.
Using this we obtain \cite{Denef:2004ze}\footnote{In general, the integral is given by $\int_S e(\nabla)$, the integral of the Euler density derived from the covariant derivative $\nabla$ \cite{Denef}.}
\be\label{counting_formula1}
I_{\rm vac} ={\rm vol}_{R_{\rm max}}({\mathbb B}^{4n+4})
 \int_S {{{\rm det} ({\cal R}+\omega\cdot {\bf 1})} \over \pi^{1+ n}},
\ee
where $S$ is now taken to be a region in $\mathcal{M} = {\cal M}_{\rm ax-dil}\times {\cal M}_{\rm cpx}(X)$, the product of the axion-dilaton moduli space and the complex structure moduli space of the three-fold  $X$. Again, $\omega$ is the K{\"a}hler form on $\mathcal{M}$, in terms of which the volume form on $\mathcal{M}$ is given by $\omega^{n + 1}/(n + 1)!$ and $\mathcal{R}$ is the Ricci curvature.

When the Ricci curvature is ignored, the geometric factor is the moduli space volume up to an overall multiplicative factor of $(n+1)!/\pi^{n+1}$:
\be\label{factor}
 \int_S {{{\rm det} (\omega\cdot {\bf 1})} \over \pi^{n+1}} = \frac{(n+1)!}{\pi^{n+1}} {\rm vol}(S).
\ee
 We briefly review the derivation of the index density, emphasizing the appearance of the Weil-Petersson metric in its canonical normalisation, following \cite{Denef}, in Appendix \ref{app:derivation}.

In this note, we will take the region $S$ to be the entire (orientifold) moduli space. As quantified in  \cite{Denef:2004ze}, using results from \cite{Gruber}, for any region $S$ in moduli space there will be corrections to the continuum-flux approximation. If $L_{\rm max}$ is large enough, then the number of lattice points in a corresponding region in flux space which contains vacua that satisfy equation \ref{eq:tadpole} will be well-approximated by the volume of that region; the leading corrections depend on the surface area of the region. When one takes $S$ to be the entire moduli space, the validity of the continuum approximation used in this note translates to the requirement that $L_{\rm max} > c\cdot b $ for some order one constant $c$. We refer to \cite{Denef:2004ze, Denef} for a more thorough discussion.

Without further input on the corresponding four-fold Euler characteristic, the maximal flux $L_{\rm max}$ is usually chosen by hand to be of order $10^{1\sim 3} $.  See \cite{Klemm:1996ts,Kreuzer:2000xy} for a list of Calabi--Yau four-folds that can be realised as hypersurfaces in toric varieties and their Euler characteristics.

%In deriving the index density $\frac{{\rm det} ({\cal R}+\omega\cdot {\bf 1})}{\pi^{n_- + 1}}$, \cite{AshokDouglas} consider an ensemble of  superpotentials $W$ drawn from a Gaussian distribution, where $W$ is the superpotential that appears in the $\mathcal{N} =1$ supergravity theory associated to a IIB flux vacuum. In their conventions, $W$ is understood to be a section of a line bundle $\cal L$ such that $c_1(\cal L) = \left[{\kappa \omega \over \pi}\right]$\footnote{For the precise mathematical definition of $\omega$, see \cite{Douglas:2004zu}.}, where $\kappa := -{1 \over M_{pl}^2}$, the four dimensional Planck mass.

%They then set $M_{pl} =1$ and compute the volumes with this choice, i.e. the natural volume form with these choices is $dV = \omega^{d}/(d)!$, and the total volume is therefore ${\pi^d \over d!}(c_1({\mathcal{L}}))^d$, with $d:= n_- + 1$ the dimension of the moduli space over which we integrate (including both complex structure moduli and the axio-dilaton). If we also elect to fix $M_{pl}=1$, then it remains to compare the numerical factors with those appearing in the canonical normalization of the Zamolodchikov metric, a task to which we turn in the next subsection. We first justify that our count of flux vacua is insensitive to arbitrary rescalings of $M_{pl}$.

The estimate for the number of flux vacua led to some effort and progress in
understanding the moduli space volume in the Weil--Petersson metric.
In particular  in  \cite{Douglas:2005hq} it was shown that the moduli space volume is finite.
\footnote{It had previously been conjectured to be finite in \cite{HorneMoore}, based on a
number of examples where it could be shown to be finite. The reason the finiteness of the
volume was important to \cite{HorneMoore} was
that a finite volume of moduli space would then lead to a well-defined probability distribution
on moduli spaces of vacua. In particular, potential energy functions generated by nonperturbative
string effects would lead to basins of attraction in moduli space. Then,
it was proposed, vacua should be selected on a statistical basis.}
However, to the best of our knowledge no moduli space volumes of Calabi--Yau three-folds leading to ${\cal N}=2, d=4$ compactifications have been computed so far. As a result, the geometric factor in the counting formula \eq{counting_formula1} is usually taken to be of order one in the estimates. In particular, the problem is often simplified to that of counting lattice points in a region in a sphere of radius $\sqrt{2 L_{\rm max}}$, whose volume accounts for the factor that should be multiplied by the `geometric' factor coming from the Calabi--Yau moduli space. See for instance \cite{Taylor:2015xtz} where this simplified estimate (i.e. neglecting the geometric factor) leads to the interesting conclusion that a single fourfold dominates the whole F-theory landscape. The contributions from other F-theory flux vacua, according to \cite{Taylor:2015xtz}, are relatively suppressed by several orders of magnitude.

The geometric factor accounts for the difference between counting fluxes that satisfy the tadpole constraint and counting
(with signs) the actual supersymmetric vacua. If the geometric factor turns out to be prodigious then we can conclude that
at least some fluxes $N$ lead to superpotentials $W_N$ admitting many vacua. (One would expect that the generic flux $N$
would lead to many vacua.) If the geometric factor turns out to be miniscule then we would be tempted to conclude that
for most flux vacua $N$, the superpotential $W_N$ in fact does \underline{not} have a supersymmetric vacuum. One cannot
arrive at this conclusion in strict logic because we are computing an index: A miniscule geometric factor might just
indicate that many vacua have cancelling contributions. Indeed, we will see an example below where the geometric
factor is negative.

In this note we compute exactly the volume of the vector multiplet moduli of type IIB compactifications on certain Calabi--Yau threefolds which lead to ${\cal N}=2, d=4$ theories before turning on the fluxes. We see that, at least in this specific family of  threefolds, it is possible that including the volume factor can lead to non-negligible effects in the counting of flux vacua. Moreover, at least for some special threefolds with non-generic holonomy (of the form $SU(2)\times G\subset SU(3)$ for some finite group $G$), we find circumstantial evidence that the volume factor decreases with increasing $b_3$, at least up to a certain critical value of $b_3$. Note that naively \eq{counting_formula1} suggests that $I_{\rm vac} $ increases with $b$ when the geometric factor is ignored \footnote{The increase of $I_{\rm vac}$ with $b$ only persists until $b/2 =  2\pi L_{\rm max}$ (recall $b/2 := 2n + 2$), after which point it decreases precipitously, as expected for the volume of a sphere of large dimension. However, as explained above, we will focus on the regime where $L_{\rm max} \gtrsim b$.}. Our result hence suggests that further study is needed to arrive at this conclusion, due to the effect of the geometric factor.

The geometric factor, of course, is not just the volume. For the special Calabi-Yaus we study, we are able to account for the Ricci curvature explicitly using the simple form of the resulting moduli spaces (Hermitian symmetric spaces). It would be nice to be able to prove something like boundedness properties of $\mathcal{R} + \omega$ on more general moduli spaces.

We also note in passing that a second application of the Calabi-Yau moduli space volumes relates to counting attractor black holes in certain string compactifications. For the counting of attractor points in type IIB compactifications, the asymptotic density of attractor points with large $|Z| \leq  Z_{\text max}$ (corresponding to a bound on the BH entropy) in a region $S$ of the complex structure moduli space is given by \cite{Denef:2004ze}
\be
{\cal N}(R,|Z| \leq Z_{\rm max})\sim {2^{n+1}\over (n+1)\pi^n} Z_{\rm max}^{n+1} {\rm vol}(S)
\ee
where ${\rm vol}(R)$ is the Weil--Petersson metric of the region $R$ and $n$ is the complex dimension of complex structure moduli space, and $n=h_{2,1}$ for a Calabi--Yau threefold.

\section{Simple volume formulas}

We now turn to the description of the volume formula for certain special Calabi-Yau moduli spaces. Often in string theory we encounter moduli spaces of string vacua that are certain double coset spaces (or products thereof) of the form
\begin{equation}
\Gamma \backslash G / K
\end{equation}
for some group $G$, (maximal) compact subgroup $K$ and discrete subgroup $\Gamma$. For example, these are familiar from the Narain moduli spaces of string compactifications on a torus $T^k$, where $G = O(k, k), K= O(k)\times O(k)$ and $\Gamma = O(k, k; \mathbb{Z})$ is the group of T-dualities. More precisely, for $L$ the underlying lattice, the group $\Gamma$ is (a subgroup of) the group of automorphisms of the lattice, which we will often denote by $O_{\mathbb{Z}}(L)$ (or, by slight abuse of notation for brevity $O(L)$). In this note, $G = O_{\mathbb{R}}(\Gamma)$ (and $K$ the maximal compact of the latter). The relatively `tame'  nature of these spaces is a consequence of some special properties of the underlying geometry, such as the preservation of extended supersymmetry; moduli spaces of $K3$ sigma models and symmetric products thereof, which possess $\mathcal{N}=4$ supersymmetry, furnish other famous examples of double coset spaces that find a natural home in string theory. In this note we aim to understand the volumes of moduli spaces  associated to Calabi-Yau manifolds that \emph{preserve only $\mathcal{N}=2$ supersymmetry but nonetheless enjoy a moduli space of double coset type}.

The moduli spaces we will be interested in are special cases of what are called Shimura varieties. We will not need the general definition of such spaces here \cite{Deligne1, Deligne2} (see \cite{Milne} for an introduction), but we note that in the special case that $G$ is of orthogonal type and signature $(2, n)$ the variety is a (quotient of a) Hermitian symmetric space and may therefore be endowed with a natural complex structure. One can go further and develop the theory of automorphic forms on such spaces, and much more. Our primary interest will be in the volume of such spaces, for which explicit formulas have happily been developed (c.f. Equation (\ref{eq:HM})); see \cite{GHS, Fiori} for further mathematical applications of these volumes, such as their appearance in (the leading term of) the growth of the dimension of spaces of cusp forms.

Here we present and explain some aspects of the formula for volumes of orthogonal Shimura varieties, following \cite{GHS, Fiori} (to which we refer the reader for further details), which build off the seminal work of Siegel \cite{Siegel}. Since several volumes appear in this note, we begin this section with a short account of the volumes and the various relationships among them. Our primary interest is in the Weil-Petersson volume ${\rm vol}_{WP}$ and we will determine the appropriate multiplicative factors to convert to ${\rm vol}_{WP}$ from the other volumes that appear in this note. The definition of the Weil-Petersson volume, and its appearance in the study of counting flux vacua, is reviewed in Appendix \ref{app:derivation}.

We first introduce the volumes computed by Siegel who computed volumes of quotients of symmetric spaces by arithmetic subgroups, ${\rm vol_S}(\Gamma \backslash {\cal D}_{rs})$ (see Equation \ref{eq:Siegelvol}). Next, we relate the Siegel volumes to the Hirzebruch-Mumford volume (Equation \ref{eq:HMvol}) employed by \cite{GHS}, which is given by a ratio of Siegel volumes: ${\rm vol}_{HM}(\Gamma \backslash \mathcal{D}_{rs}):= {{\rm vol_S}(\Gamma \backslash \mathcal{D}_{rs}) \over {\rm vol_S}(\mathcal{D}^{(c)}_{rs})}$, where $\mathcal{D}^{(c)}_{rs}$ is the compact dual of $\mathcal{D}_{rs}$, given below. This is a natural volume from a mathematical perspective and, since we closely follow the presentation of \cite{GHS}, we take time to introduce it. We also use several computations of ${\rm vol}_{HM}$ in \cite{GHS} for interesting classes of spaces, and convert them to computations of ${\rm vol}_{WP}$, in Section \ref{sec:estimates}.

The Hirzebruch-Mumford volume may be immediately compared to both the canonically normalised Zamoldchikov volume, familiar to physicists, and computed for several interesting classes of spaces in \cite{Moore}, as well as to the canonically normalised Weil-Petersson volume. The conversion factor between the Weil-Petersson and Hirzebruch-Mumford volumes appears in  Equation \ref{eq:norm}. We also fix the relative normalisations between the Weil-Petersson and Zamoldchikov volumes in Appendix \ref{app:norm} using standard string theoretic manipulations.

We begin with the Siegel volumes. Siegel began his study with the homogeneous symmetric domain
\begin{equation}\label{hsd}\mathcal{D}_{rs} = O(r, s)/ O(r) \times O(s)\end{equation} using its realization as a bounded domain:
\begin{equation}
\mathcal{D}_{rs} = \left\lbrace X \in \text{Mat}_{\,r \times s}(\mathbb{R})| I_r - X X^t >0 \right\rbrace.
\end{equation}
This proceeds by making use of the natural $O(r, s)$-invariant metric
\begin{equation}
ds^2 = {\rm Tr}\left((I_r - X X^t)^{-1} dX (I_s - X^t X)^{-1}d X^{t} \right)
\end{equation} which induces the following volume form on $\mathcal{D}_{rs}$:
\begin{equation}
dV = \left({\rm det}(I_r - X X^t)^{-1} \right)^{r + s\over2} \prod_{i, j}dx_{ij}.
\end{equation}
With respect to this volume form, Siegel then computes, for any lattice of signature $(r,s)$:
\begin{equation}\label{eq:Siegelvol}
{\rm vol}_S(O(L) \backslash \mathcal{D}_{rs}) = 2 \alpha_{\infty}(L) |{\rm det} L |^{(r + s + 1)/2}\gamma_r^{-1} \gamma_s^{-1}
\end{equation}
where $\gamma_m := \prod_{k=1}^m \pi^{k/2}\Gamma(k/2)^{-1}$, $\alpha_{\infty}(L)$ is the real Haar measure of $L$, also known as the Tamagawa measure, and ${\rm det} L$ is simply the determinant of the matrix whose $(ij)$th entry is the inner product of the $i$th and $j$th basis vector with respect to a chosen basis\footnote{This matrix is often called the Gram matrix.}. We will refer to the above volume as the {\it Siegel volume}.

Next, we decompose the Lie algebra $\mathfrak{g}$ of $O(r, s)$ as $\mathfrak{g} = \mathfrak{k}\oplus\mathfrak{p}$, where $\mathfrak{k}$ is the Lie algebra of $O(r)\times O(s)$ and $\mathfrak{p}$ is the orthogonal complement with respect to the Killing form and may be written as \begin{equation}\mathfrak{p} = \left\lbrace \begin{pmatrix} 0 & U \\^tU & 0  \end{pmatrix}, U \in {\rm Mat}_{r\times s}(\mathbb{R})\right\rbrace.\end{equation} This parabolic subspace is isomorphic to the tangent space of $\mathcal{D}_{rs}$ at the origin. Indeed, starting from the Killing form $tr(U_1 ^tU_2)$ one may produce the $O(r, s)$-invariant metric by studying the tangent space at the origin. We also introduce the compact dual of our symmetric space: $\mathcal{D}^{(c)}_{rs} = SO(r + s)/SO(r) \times SO(s)$.\footnote{Note that $O(r, s)/O(r)\times O(s) = SO(r, s)_{0}/SO(r) \times SO(s)$, where $SO(r, s)_0$ is the component connected to the identity.} The tangent space of $\mathcal{D}^{(c)}_{rs}$ at the identity $I_{r+s}$ is given by
\begin{equation}
\mathfrak{p}' = \left\lbrace \begin{pmatrix} 0 & U \\-^tU & 0  \end{pmatrix}, U \in {\rm Mat}_{r\times s}(\mathbb{R})\right\rbrace,
\end{equation} and the Killing form of $SO(r+s)$ induces the form $2 {\rm Tr}(U_1^tU_2)$ on $\mathfrak{p}'$ \cite{Helgason}. To properly compare the volumes of $\Gamma \backslash \mathcal{D}_{rs}$ and $\mathcal{D}^{(c)}_{rs}$, as required to produce the Hirzebruch-Mumford volume, one has to normalise the metrics on $\mathcal{D}_{rs}$ and $\mathcal{D}^{(c)}_{rs}$ so that they coincide with the Siegel metric at their common base point. For instance, when computing the volume of $SO(r+s)$ using the metric induced from the Killing form one must multiply by an additional factor of $2^{-(r+s)(r + s-1)/4}$, using the fact that the dimension of $SO(n)$ is $n(n-1)/2$ \cite{GHS}. In total, the Siegel volume of the compact dual is ${\rm vol}_S(\mathcal{D}^{(c)}_{rs}) = 2 \gamma_{r + s} \gamma_r^{-1} \gamma_s^{-1}$.

To finish the computation of the volume, we still need to determine the Tamagawa measure $\alpha_{\infty}(L)$. It turns out \cite{GHS, Fiori} that the Tamagawa measure may be computed in terms of local densities of lattices $L\otimes \mathbb{Z}_p$ over the $p$-adic integers:
\begin{equation}
\alpha_{\infty}(L) = {2 \over g^+_{sp}(L)}\prod_p \alpha_p(L)^{-1},
\end{equation}
where $g^+_{sp}(L)$ is the number of proper spinor genera in the genus of $L$. Importantly for us, the right hand side is computable for a given lattice $L$. We record the definitions of the proper spinor genera, and local factors $\alpha_p(L)$,  in Appendix \ref{app:padic}, and refer to \cite{Fiori} for the algorithm with which one may compute them.

At last, the Hirzebruch-Mumford volume as determined by \cite{GHS, Fiori} in the notation of \cite{GHS} is given by
\begin{equation}\label{eq:HMvol}
{\rm vol}_{\rm HM}(O(L)\backslash \mathcal{D}_{rs}) = {2 \over g^+_{sp}(L)} |{\rm det} L|^{(r + s + 1)/2} \prod_{k=1}^{r + s}\pi^{-k/2} \Gamma(k/2) \prod_p \alpha_p(L)^{-1}.
\end{equation}
When $L$ is a lattice of signature $(2, n), n\geq 1$ and contains at least one hyperbolic plane (the primary case of interest for us), then the formula specializes to \cite{GHS}
\begin{equation}
{\rm vol}_{\rm HM}(O(L)\backslash \mathcal{D}_{rs}) = 2 |{\rm det} L|^{(n + 3)/2}\prod_{k=1}^{n+2}\pi^{-k/2}\Gamma(k/2)\prod_p \alpha_p(L)^{-1}.
\end{equation}
This specialization uses the facts that a.) the spinor genus of an indefinite lattice of rank $\geq 3$ coincides with its class and b.) the genus of any indefinite lattice containing a hyperbolic plane contains only one class \cite{Kneser}.
If one wishes to study the volume with respect to a choice of finite index arithmetic subgroup of $O(L)$, which we denote by $\Gamma$, and if we still focus on $L$ of signature $(2, n)$ and containing a hyperbolic plane, the Hirzebruch-Mumford volume is given by
\begin{equation}
{\rm vol}_{\rm HM}(\Gamma\backslash \mathcal{D}_{rs}) = 2 \left[PO(L):P\Gamma \right] |{\rm det} L|^{(n + 3)/2}\prod_{k=1}^{n+2}\pi^{-k/2}\Gamma(k/2)\prod_p \alpha_p(L)^{-1},
\end{equation}
where the notation $PG$ refers to the image of the group in ${\rm Aut}(\mathcal{D}_{rs})$ (which is isomorphic to the group modulo its center).

In what follows, we will sometimes denote volumes  ${\rm vol}(\Gamma\backslash \mathcal{D}_{rs})$ by simply ${\rm vol}(\Gamma)$ or ${\rm vol}(O(\Gamma))$, with the understanding that we are always computing volumes of double coset spaces.

Next, we will determine the factor that converts the Hirzebruch-Mumford volume to the (canonically normalised) Weil-Petersson metric for our physical applications. To do this, we will first compare the Hirzebruch-Mumford volume to the Zamolodchikov volume studied in \cite{Moore} as an intermediate step. In Appendix \ref{app:norm}, we will compute the conversion factor between the Zamolodchikov and Weil-Petersson metrics. Combining these contributions, we will presently obtain
\begin{equation}\label{relative_factor}
{\rm vol}_{\rm WP}(\Gamma) = C_{\rm WP}{\rm vol}_{\rm HM}(\Gamma)
\end{equation}
where
\begin{equation}\label{eq:norm}
C_{\rm WP}= \left(1 \over \sqrt{2} \right)^{2n}\frac{\sigma(2 + n)}{\sigma(2)\sigma(n)}
\end{equation}
with $\sigma(D) \equiv 2^{(D + 1)/2}\prod_{j=1}^{D-1}\left({(2 \pi)^{{j + 1 \over 2}} \over \Gamma({j + 1 \over 2})} \right)$.

We first recapitulate the Zamolodchikov volumes computed in \cite{Moore}. Consider first the double coset
\be
\mathcal{N}_{a + 8b, a} = O_{\mathbb{Z}}(Q_{a, b})\backslash O_{\mathbb{R}}(Q_{a, b})/(O(a + 8 b)\times O(a))
\ee
where $Q_{a, b}$ denotes the quadratic form of the \textit{even, unimodular} lattices of signature $(a + 8 b, a)$.
Its volume, induced from the natural left-right invariant metric on the Lie algebra, is given by
\be \label{eq:vol1}
\textrm{vol}^{tr}(\mathcal{N}_{a + 8b, a}) = \frac{\sigma(2 a + 8b)}{\sigma(a)\sigma(a + 8b)}2(d-1)!\frac{\zeta(d)}{(2 \pi)^d}\prod_{j=1}^{d-1}\frac{|B_{2j}|}{4 j},
\ee
with $\sigma(D)$ defined as above, and the Zamolodchikov metric is
\begin{equation}\label{eq:volZamo}
{\rm vol}_{Z}(\mathcal{N}_{a + 8b, a}) = \left({1 \over \sqrt{2} \pi}\right)^{a(a + 8b)}{\rm vol}^{tr}(\mathcal{N}_{a + 8b, a}).
\end{equation}
If we specialize this result to even, unimodular lattices of signature $(2, 2 + 8b)$, we can compute the Zamolodchikov/HM conversion directly to be
\begin{equation}
{\rm vol}_{Z}(\mathcal{N}_{2 + 8b, 2}) = \left({1 \over \sqrt{2} \pi}\right)^{2(2 + 8b)}\frac{\sigma(4 + 8b)}{\sigma(2)\sigma(2+8b)}{\rm vol}_{HM}(\mathcal{N}_{2 + 8b, 2}).
\end{equation}

More generally, it is derived in \cite{Moore} that $ds_{Z}^2 = {1 \over 2 \pi^2} ds^{2, tr}$. Notice that for a lattice of signature $(2, n)$, $\frac{\sigma(2 + n)}{\sigma(2)\sigma(n)} = 2^{n} {\gamma_{2 + n} \over \gamma_2 \gamma_n}$, so that, up to the factors of $\pi$, the conversion is essentially reinstating the volume of the compact dual that is divided out in the definition of ${\rm vol}_{HM}$ \footnote{To account for the various factors of 2 that arise in the conversion see the discussion above and \cite{GHS} and \cite{Moore}}.

Finally, we derive in Appendix \ref{app:norm} that $ds^2_{WP} = \pi^2 ds^2_Z$, which leads to the relation \ref{relative_factor}.
%Orthogonal Shimura Varieties

\section{Compactification on the Enriques Calabi-Yau}\label{sec:FHSV}

The FHSV model \cite{FHSV} is a particularly simple example of a compactification down to four dimensions that preserves $\mathcal{N}=2$ spacetime supersymmetry. First, let us briefly recall its  basic properties. Our presentation will largely follow \cite{FHSV, Aspinwall}. The FHSV model is obtained via string theory compactification on the so-called Enriques Calabi-Yau manifold, which is a quotient of $K3 \times T^2$ by a fixed-point-free involution. More specifically, one considers a free Enriques involution on the $K3$ factor but allows the involution on the  $T^2$ to have fixed points. Consequently, the theory possesses $\mathcal{N}=2$ spacetime supersymmetry but in some aspects enjoys similar physics to the underlying $\mathcal{N}=4$ theory. One avatar of this is that the Enriques Calabi-Yau manifold has $SU(2) \times \mathbb{Z}_2$ holonomy, rather than $SU(3)$ holonomy. In particular, the involution just acts as $-1$ on the torus coordinate $z_3$, and as $-1$ on the holomorphic $(2, 0)$-form $\Omega$ on the $K3$, giving a natural invariant $(3, 0)$-form $\Omega \wedge dz_3$. The resulting Enriques surface has nonvanishing Hodge number $h^{(1, 1)}=10$ and the  full Enriques Calabi-Yau has $h^{(1, 1)}(X_3) = 11$. One can also compute that the manifold has $h^{(2, 1)}(X_3) = 11$, and is self-mirror up to a global $\mathbb{Z}_2$ discrete torsion. The latter implies that instanton corrections vanish in this model, meaning the classical moduli spaces, described below, are in fact locally exact.

%For a type IIB compactification on an Enriques Calabi-Yau manifold,
The complex structure moduli space of the Enrique Calabi-Yau (which in IIB compactification is part of the vector multiplet moduli space) takes the form
\be\label{eq:vectors}
\left(SL(2, \mathbb{Z})\backslash SL(2, \mathbb{R})/ SO(2) \right) \times \left(O(\Gamma^{2, 10})\backslash O(2, 10)/(O(2)\times O(10)) \right).
\ee The first factor arises from the complex modulus of a complex torus and the second factor from an Enriques surface.
In the second factor, we have
\be\label{FHSVlattice}
\Gamma^{2, 10} : = \Gamma^{1, 1}\oplus \Gamma^{1, 1}(2)\oplus E_8(-2).
\ee

%We leave off the contribution of the axio-dilaton, i.e. another copy of the fundamental domain, for now.
Notice that the perturbative in  $\alpha'$ correction to the prepotential (of order $\alpha'^3$) vanishes for this Calabi-Yau because the term is proportional to its Euler characteristic, $\chi = 2(h^{1, 1} - h^{2, 1})= 2(11-11)=0$\cite{Grimm}. In addition, the first, genus zero, non-perturbative corrections to the prepotential vanish as well.

\subsection{Moduli space volume}
We will now compute the volumes of the full complex structure moduli space and the orientifold moduli space. We can directly compute the volume by starting with Equation \ref{eq:HMvol}, computing the local densities and other lattice-dependent contributions, and converting it to the Weil-Petersson normalisation using Equation \ref{relative_factor}.

Our FHSV lattice $\Gamma^{2, 10}$ is very close to the unimodular lattice $2 \Gamma^{1, 1}\oplus E_8(-1)$ and we will show that its volume differs from its unimodular counterpart by an overall factor $2079/2 \sim 10^3$, by recomputing the appropriate local densities and determinant factor. First, we get a contribution of $(2^{10})^{13/2}$ from the factor $|{\rm det} \Gamma^{2, 10}|^{(r + s + 1)/2}$. Additionally, relative to the unimodular case, the rescaling of the constituent sublattices will change the contribution coming from the local factor $\alpha_2(\Gamma^{2, 10})^{-1}$, but none of the other factors.

To compute local densities one should know the Jordan decomposition of the lattice $\Gamma^{2, 10}$ over $\mathbb{Z}_p$, the $p$-adic integers; see Appendix \ref{app:padic} for the definition of the Jordan decomposition and several examples. We can express a so-called $p^r$-modular lattice $L$ as the appropriate rescaling of a unimodular lattice $N$, $N(p^r)$, and we will be interested in the decomposition of a general lattice $L$ into $p^j$-modular lattices $L_j$ of ranks $n_j$ which are $p^j$-rescalings of unimodular lattices $N_j$. In equations, $L = \bigoplus_{j \in \mathbb{Z}}L_j$ where $L_j:= N_j(p^j)$. With this notation, the local density of interest is given by (see \cite{GHS} for the most general definition of these quantities, and for notation; below we already make several simplifications for our lattice of interest)
\begin{equation}
\alpha_2(L) = 2^{n - 1 + w} P_2(L) E_2(L)
\end{equation}
with
\begin{gather}
\begin{split}
w &= \sum_{j} j n_j \left(\frac{n_j + 1}{2} + \sum_{k>j} n_k \right)\\
P_2(L) &= \prod_{j} P_2\left({{\rm rank}(N_j)\over2}\right)~~{\rm with}~~P_2(n) = \prod_{i=1}^n (1 - 2^{-2i})\\
E_2(L) &=\prod_{j, L_j \neq 0}\frac{2}{1 + 2^{-{{\rm rank}(N_j)/2}}}.
\end{split}
\end{gather}

The Jordan decomposition for our lattice over $p\neq2$ is given by $\Gamma^{2, 10}\otimes \mathbb{Z}_p = 6\Gamma^{1, 1}$, so the local densities for $p \neq 2$ coincide for those of the unimodular lattice of signature $(2, 10)$ and are given in \cite{GHS}. The decomposition for $\Gamma^{2, 10}$ over $\mathbb{Z}_2$, on the other hand, is given by $5 \Gamma^{1,1}(2) \oplus \Gamma^{1,1}$, which is the sum of five $2^1$-modular lattices and one unimodular lattice. The corresponding local density is the only thing we need to compute, and plugging everything in to the previous definitions we find $w = 55, \alpha_2(\Gamma^{2, 10}) = 98563190995235635200$, and therefore an overall discrepancy, including the determinant factor, of ${2079 \over 2}$ from the unimodular lattice of the same signature.

If we plug in $(a=2, b=1)$ to (\ref{eq:vol1}) and multiply by our compensatory factor we get
\be \label{WPfund}
\textrm{vol}^{tr}_{\rm vec1}(\Gamma^{2, 10})=  \frac{\pi ^{10}}{320820302880000}{2079 \over 2} \sim 3 \times 10^{-7}.
\ee
The subscript indicates that this is the volume of one factor of the full vector multiplet moduli space.

Next, we need the volume for the first factor of the vector multiplet moduli space (\ref{eq:vectors}), which is the familiar modular fundamental domain of the upper half-plane. The volume computed in the standard Poincar{\'e} metric (writing $\tau = x + i y$) is well-known to be
\be
\int_{\cal F} {dx\, dy \over y^2} = {\pi \over 3}.
\ee
Applying our previous formulas, the volume of the fundamental domain with respect to the Weil-Petersson metric is given by ${1 \over 2}{\rm vol}^{tr}({\cal F}) = {1 \over 2}{\pi \over 6}$:
\be
\int_{\cal F} {dx\, dy \over 4 y^2} = {\pi \over 12}.
\ee
Note that, as a consistency check, our normalisation gives the same volume of the fundamental domain as that computed in \cite{AshokDouglas}.

%Below and in Appendix \ref{appA}, we will fix the overall normalization of the volume that is canonical for the Weil-Petersson metric; notice that in these conventions we will reproduce the volume computed by \cite{AshokDouglas}.

Putting together the Weil-Petersson-normalised volumes for both factors, we obtain
\begin{gather}
\label{eq:volFHSV}
\begin{split}
\textrm{vol}_{\rm WP}(\mathcal{M}_{\rm cpx})& = \left(\left({1 \over \sqrt{2}}\right)^2 {\pi \over 6}\times \left({1 \over \sqrt{2}}\right)^{20} \textrm{vol}^{tr}_{\rm vec1}(2, 1)\right) \\ &
= \frac{\pi ^{11}}{3792438558720000} \sim 7.7\times 10^{-11}.
\end{split}
\end{gather}

Notice that the axio-dilaton moduli space computed with respect to this metric contributes an additional factor ${\rm vol}_{\rm WP}(\mathcal{M}_{\rm ax-dil}) = \pi/12$ as well (cf. \eq{WPfund}):

\begin{equation}
\textrm{vol}_{\rm WP}(\mathcal{M}_r \times \mathcal{M}_{\rm axio-dil})= \textrm{vol}_{\rm WP}(\mathcal{M}_{\rm cpx})\times\textrm{vol}_{\rm WP}(\mathcal{M}_{\rm ax-dil}) = {\pi^{12} \over 45509262704640000} \sim 2.0\times10^{-11}.
\end{equation}

\subsection{Orientifold counting}

The orientifold procedure projects out some complex structure moduli, thereby reducing the dimensionality of the complex structure moduli space. Consequently, the volume taken over the whole moduli space may not be a good approximation to the volume of the remaining moduli space that the fluxes are allowed to occupy after orientifolding. Indeed, in our particular example, we will presently see that this is the case.

We will use the orientifold action studied in \cite{Grimm}. We emphasize here that we are making a particular, tractable choice of orientifold action; other choices of orientifold action may preserve more complex structure moduli and potentially result in surviving moduli spaces that are symmetric spaces for $O(2, n), n < 10$.  The involution of \cite{Grimm}, as characterized by its action on cohomology, is chosen to act as an overall minus sign on the $E_8$ lattice factor while leaving the $\Gamma^{1, 1}$ factor coming from the parent K3 surface invariant, hence acting by an overall minus sign on the Enriques surface's top form. It also acts by a minus sign on the coordinate of the $T^2/\mathbb{Z}_2$ factor. This action restricts the complex structure moduli space to a certain sublocus that has, happily, already been explored in the context of studying simplifications of the topological string on the Enriques Calabi-Yau \cite{KlemmMarino, GrimmKlemm}. Blowing down the 8 specified cycles results in the reduced moduli space $\mathcal{M}_r$ of the following local form (suppressing the axio-dilaton factor, which is untouched by the orientifold)
\begin{equation}\label{red_before_quotient}
%\mathcal{M}_r =
\frac{SL(2, \mathbb{R})}{SO(2)} \times \left(\frac{SL(2, \mathbb{R})}{SO(2)}\right)^2.
\end{equation}
The first factor, which descends from the torus, is quotiented by the discrete group $SL(2, \mathbb{Z})$ as usual, while the second factor is quotiented by the discrete group $\Gamma(2)\times \Gamma(2)$, which is deduced in \cite{KlemmMarino} by a subtle analysis.
 This form of the moduli space follows from noticing (as verified by detailed computations in \cite{KlemmMarino}) that the reduced moduli space has an algebraic realization as a product of $\Gamma(2)$-symmetric elliptic curves:
\begin{equation}
x_1^2 = x_2^4 + x_3^4 + z^{-1/4}x_1 x_2 x_3.
\end{equation}

The volume of the orientifold moduli space, being merely a product of several quotients of the upper half-plane, is obtained easily. As described earlier, the two factors in \eq{red_before_quotient} on the right have a discrete symmetry group $\Gamma(2)$, a congruence subgroup of $SL(2, \mathbb{Z})$ of index 6. Therefore, the volume of the orientifold complex structure moduli space is
\begin{equation}
\textrm{vol}_{\rm WP}(\mathcal{M}_r) = {\pi \over 12}\times \left({\pi \over 2}\right)^2
\end{equation}
which, including another ${\pi \over 12}$ from the axio-dilaton, gives $\textrm{vol}_{\rm WP}(\mathcal{M}_r \times \mathcal{M}_{\rm axio-dil}) = {\pi^4 \over 576} \sim 0.2$: an order 1 volume after all! Recall also that although the volume is of order 1, the index density is the Weil-Petersson volume multiplied by a $(n_- + 1)!/(\pi)^{n_- + 1}$ factor, and the latter brings the order of magnitudes down slightly, as we will now compute.

%\subsection{Voluminous recounting}
 With the moduli space volumes in hand, we can now ask about their (rough) quantitative impact on the statistical formulas reviewed in Section \ref{intro}, subject to the assumptions described therein.
The maximum number of  flux quanta allowed by tadpole cancellation is determined by the Euler characteristic of the Calabi-Yau fourfold $X_4$ coming from the F-theory lift (divided by 24).
For instance, in the weakly coupled limit and given an orientifold action like that in \cite{Grimm} we can glean some information about $\chi(X_4)$ in terms of the Hodge numbers of the threefold $X_3$ \cite{Denef,Klemm:1996ts}:
%Though the F-theory lift of the FHSV model is not known, we can glean some information about $\chi(X_4)$ in terms of the Hodge numbers of the threefold $X_3$ \cite{Denef}:
\begin{gather}
 \begin{split}
\chi(X_4) &= 48 + 6(h^{(1, 1)}(X_4) + h^{(3,1)}(X_4)- h^{(2, 1)}(X_4)) \\
h^{(1, 1)}(X_4)  &=  h^{(1,1)}_+(X_3)+1 \\
h^{(2,1)}(X_4) &= h^{(1, 1)}_-(X_3) \\
h^{(3, 1)}(X_4) &= h^{(2, 1)}_-(X_3) + 1 + h^{(2, 0)}(S)
\end{split}
\end{gather}
where the $\pm$ subscripts denote the eigenvalues under the orientifold action,  and $S$ denotes the surface in $X_3$ wrapped by D7 branes. Unfortunately, computing its contribution in the perturbative IIB picture is quite subtle \cite{Collinucci:2008pf} and often yields the lion's share contribution to $\chi(X_4)$.
%\textcolor{red}{MC: I've deleted the sentence  "To define a IIB flux vacuum on a Calabi-Yau threefold, we must first define an appropriate orientifold action."
%is it okay?
 %}

For instance, using the orientifold action chosen in \cite{Grimm} we immediately see that $$h^{(1, 1)}_-(X_3) = 8, ~h^{(1, 1)}_+(X_3) = 3, ~h^{(2, 1)}_+(X_3) = 8, ~h^{(2, 1)}_{-}= 3$$ which gives us the lower bound $\chi(X_4) \geq 48 + 6(4 + 4 - 8) = 48$ and hence $L_{\rm max} = {\chi(X_4)\over 24}> 2$.
Again, since $h^{(2, 0)}(S)$ is normally the dominant contribution to $\chi(X_4)$, we expect this to be a  weak lower bound. Furthermore,
the continuous approximation formulas of \cite{AshokDouglas} is strictly speaking not valid when $L_{\rm max}$  is of order one.
%an order one $L_{\rm max}$ too small to justify using the continuous approximation formulas of \cite{AshokDouglas}.
With these points in mind, we will remain somewhat agnostic about the correct value of $L_{\rm max}$\footnote{A reasonable approximation to $L_{\rm max}$ in this model, without constructing an explicit F-theory lift, may be to take the fourfold to be $K3 \times K3$, which gives $L_{\rm max} = 24$; we thank Thomas Grimm for this suggestion.} and test several values, $L_{\rm max}\sim 10^1,10^2, 10^3$, representative of contributions from `typical' fourfolds\footnote{It might be interesting to consult the lists of Hodge numbers of Calabi-Yau fourfolds represented as hypersurfaces in toric varieties. See for example \cite{Klemm:1996ts, Kreuzer:2000xy}.}. For convenience, we reproduce the formula of \cite{AshokDouglas}
\be\label{count_flux_vac}
{\cal I}_{\rm vac}(R,L \leq L_{\text{max}})\sim {(2\pi)^{2n_- + 2}\over (2n_-+2)!} L_{\text{max}}^{2n_-+2} \int_M {{\rm det}({\cal R}+\omega\cdot {\bf 1}) \over \pi^{n_-+1}},
\ee
In our example $n := h^{(2, 1)}(X_3) = 11$, we have $n_- = h^{(2, 1)}_-(X_3) = 3$, where the subscript again denotes the anti-invariant part under the orientifold involution.
\begin{table}
\begin{tabular}{ l | c | r }
  \toprule			
  $L_{\rm max}$ & ${\rm vol}_{R_{\rm max}}({\mathbb B}^{4n+4}) $ & $I_{\rm vac, vol}$ \\\midrule
  $10^1$ & $10^{16}$ & $10^{8}$ \\
  $10^2$ & $10^{34}$ & $10^{26}$ \\
  $10^3$ & $10^{52}$ & $10^{44}$ \\
  \bottomrule
\end{tabular}
\caption{Estimates of the number of flux vacua in the FHSV model. Here ${\rm vol}_{R_{\rm max}}({\mathbb B}^{4n+4})$ denotes the estimate from the $(4 h_{2, 1} + 4) = 48$-ball volume factor, assuming the moduli space volume contribution is of order 1, which has been the strategy employed in the literature so far. $I_{\rm vac, vol}$ includes the geometric contribution ${(12)! {\rm vol}_{\rm WP}(\mathcal{M}_{\rm cpx}){\rm vol}_{\rm WP}(\mathcal{M}_{\rm ax-dil}) \over \pi^{12}}$.}\label{tbl:FHSV}
\end{table}
The estimates for the number of flux vacua at various $L_{\rm max}$, assuming the volume factor is order 1, as well as accounting for the contribution of ${\rm vol}_{\rm WP}({\cal M})_{\rm cpx}$ using (\ref{eq:volFHSV}) are recorded in Table \ref{tbl:FHSV}.
%Of course, since the formula is asymptotic we should view a small number not as proof that there are few flux vacua but rather as evidence that Equation \ref{count_flux_vac} breaks down.
Of course, since the formula is asymptotic we should take the result obtained by applying \eq{count_flux_vac} with a grain of salt when it is of order one.

Accounting for the orientifold action, we also re-compute the quantities of Table \ref{tbl:FHSV} using the volumes of the orientifold sublocus and replacing $n \rightarrow n_-$; see Table \ref{tbl:orientifold}. We stress again that there may be other choices of orientifold action such that $n_- \sim n$, in which case the estimates of Table \ref{tbl:FHSV} would be more indicative of the volume factor corrections.

\begin{table}
\begin{tabular}{ l | c | r }
  \toprule			
  $L_{\rm max}$ & ${\rm vol}_{R_{\rm max}}({\mathbb B}^{4n_-+4}) $ & $I_{\rm vac, orient}$ \\\midrule
  $10^1$ & $10^{10}$ & $10^{8}$ \\
  $10^2$ & $10^{18}$ & $10^{16}$ \\
  $10^3$ & $10^{26}$ & $10^{24}$ \\
  \bottomrule
\end{tabular}
\caption{Estimates of the number of flux vacua in the FHSV model with a specific choice of orientifold action. Here ${\rm vol}_{R_{\rm max}}({\mathbb B}^{4n_-+4})$ denotes the estimate from the $(4 h^-_{2, 1} + 4) = 16$-ball volume factor, assuming the moduli space volume contribution is of order 1. $I_{\rm vac, orient}$ includes the geometric contribution ${(4)! {\rm vol}_{\rm WP}(\mathcal{M}_{r}){\rm vol}_{\rm WP}(\mathcal{M}_{ax-dil}) \over \pi^{4}}$. Note that the relatively small effect of the geometric contribution can be traced to the fact that
$n_-  = 3$ in this case, in contrast to $n = 11$ before the orientifold.}\label{tbl:orientifold}
\end{table}

\subsection{The effect of the curvature}

We now reinstate the Ricci curvature into the geometric factor in the counting formula \eq{counting_formula1}, so that we are computing the integral of an Euler density of a connection on $TS\otimes \mathcal{L}$ (see Appendix \ref{app:derivation} for the derivation), rather than the volume form. Slightly more explicitly, we have
\begin{align}
{1 \over \pi^{n+1}}{\rm det}(\mathcal{R} + \omega\cdot{\bf 1}) &={1 \over \pi^{n+1}}{\rm det}\left(R^{l}_{i \bar{j} k} dz^i \wedge d z^{\bar j} +  \delta^l_k{i \over 2}g_{i \bar{j}}dz^i \wedge dz^{\bar{j}} \right)
\end{align} where the curvature two-form is expressed as a Hermitian $(n+1)\times(n+1)$ matrix and is given in terms of the Hermitian metric $g_{i \bar{j}}$ as $R^{l}_{i \bar{j} k} = -{\rm i} g^{l \bar{m}}R_{i \bar{j} k \bar{m}}$.

The computation in the index density is particularly simple in the case of the FHSV orientifold. The curvature matrix decomposes into three $1\times1$ matrices which we denote by ${\cal R}_{0, 1, 2}$ and each of the three upper half plane enjoys the relation $\mathcal{R}_a = - 2 \omega_a$. Explicitly, we have:
\begin{gather}
 \begin{split}
{\textrm{det}(\mathcal{R} + \omega\cdot\mathbf{1}) \over \pi^3} &= {1 \over \pi^3}\,\textrm{det}\begin{pmatrix}
\mathcal{R}_1 + \sum_{i=1}^3 \omega_i & 0 & 0 \\
0 & \mathcal{R}_2 + \sum_{i=1}^3 \omega_i & 0 \\
0& 0 & \mathcal{R}_3 + \sum_{i=1}^3 \omega_i
\end{pmatrix} \\
&= {1 \over \pi^3}\,\textrm{det}\begin{pmatrix}
-\omega_1 + \omega_2 + \omega_3 & 0 & 0 \\
0 & \omega_1 - \omega_2 + \omega_3  & 0 \\
0& 0 & \omega_1 + \omega_2 - \omega_3
\end{pmatrix} \\
&= {-2 \over \pi^3}(\omega_1 \wedge \omega_2 \wedge \omega_3)
\end{split}
\end{gather}
and so the curvature contribution has modified the answer by $-2$.

More generally, if   $(X,g)$ is an Hermitian symmetric space of real dimension $d=2n$ then,
giving $X$ a natural complex structure from a choice of positive roots, and letting
 $R$ denote the curvature 2-form of the holomorphic tangent space and $\omega$ the K\"ahler form, we claim that
\be
\det (R + \omega) =   (1  -  \frac{2}{d-1} )\omega^n
\ee
To prove this note that
\be
R_{\mu\nu\lambda\rho} = \kappa ( g_{\mu\lambda} g_{\nu\rho} - g_{\mu \rho} g_{\nu \lambda})
\ee
with $\kappa = \frac{2}{d-1}$.
\footnote{To check the normalisation we compute the Ricci tensor and refer to Proposition 3.6, ch. VIII
of \cite{Helgason}.
}
We can choose local coordinates so that
the metric is
\be
ds^2 = g_{\mu\nu} dx^{\mu} \otimes dx^{\nu} = \sum_{i=1}^n \lambda_i ( (du^i)^2 + (dv^i)^2)
\ee
with complex coordinates
\be
\begin{split}
z^j & = u^j + {\rm i} v^j \\
\bar z^j & = u^j - {\rm i} v^j \\
\end{split}
\ee
In these coordinates the curvature $2$-form is an outer product of two vectors: 
\be
R^i_{~j} = - \frac{{\rm i}\kappa}{2} dz^i \lambda_j d \bar z^{\bar j} ~ . 
\ee
We now use the identity
\be\label{eq:easy}
\det( x \delta_{ij}  + v_i w_j ) = x^n + x^{n-1} \left( \sum_i v_i w_i \right)
\ee
which holds over an arbitrary commutative ring.

The generalization to a product of symmetric spaces is straightforward.
Now $R^i_{~j}$ is a block diagonal matrix. Consider for instance a produce of two symmetric spaces.  Letting $\kappa_1 , \kappa_2$
be the constants for the two factors, with K\"ahler forms $\omega_1, \omega_2$, and so forth,
we have
\begin{gather}
\begin{split}
\det (R + \omega) & =   \left[ (\omega_1 + \omega_2)^{n_1} - \kappa_1 (\omega_1 + \omega_2)^{n_1-1} \omega_1 \right]
\cdot
\left[ (\omega_1 + \omega_2)^{n_2} - \kappa_2 (\omega_1 + \omega_2)^{n_2-1} \omega_2 \right] \\
& = \omega^{n-2} \left[ \omega^2 - \kappa_1 \omega_1^2  -\kappa_2  \omega_2^2  -(\kappa_1 + \kappa_2 - \kappa_1\kappa_2) \omega_1 \omega_2
\right] \\
\end{split}
\end{gather}
where $\omega = \omega_1 + \omega_2$. We conclude that for products of homogeneous spaces, the
inclusion of the two form $R$ in the geometric factor does not produce a significant difference
from the volume.

%we obtain an order one number from the counting formula \eq{\ref{count_flux_vac}}. }

%\subsection{Curvature in the orientifold}

\section{Estimation of other models}\label{sec:estimates}

In this section, we conduct a similar analysis on related models, and discuss possible lessons one can learn for a more general class of string compactifications.

\subsection{Generalities in signature $(2, n)$}
Roughly speaking, lattices of signature $(2, n)$ comprise a family of examples with computable moduli space volumes that moreover are relevant in string compactifications. We reproduce for convenience the general formula for the Hirzebruch-Mumford volume for a lattice of signature of $(2, n)$, i.e. the volume of the double coset moduli space $O(L) \backslash O(2, n) / O(2) \times O(n)$ \cite{GHS, Fiori}:
\begin{equation}\label{eq:HM}
{\rm vol}_{\rm HM}(O(L)) = \frac{2}{g^+_{sp}} |{\rm det} L|^{(n+2 + 1)/2}\prod_{k=1}^{n+2}\pi^{-k/2}\Gamma(k/2)\prod_p \alpha_p(L)^{-1}
\end{equation}
where %$\rho \equiv 2 + n$ is the rank of the lattice,
$g^+_{sp}$ is the number of proper spinor genera in the genus of $L$ and $\alpha_p(L)$ are the local factors. We will also include, at the end, our overall normalisation factor to put the Hirzebruch-Mumford volumes in the canonical normalisation of the Weil-Petersson metric. Fixing all the constants and computing the local factors in the formula \eq{eq:HM} requires specifying some details of the class of lattices under consideration (e.g. Are the lattices even and unimodular? Do they contain some factors of the standard hyperbolic lattice $\Gamma^{1, 1}$?) In this section we will try to study some general expectations for how the volumes scale with $n$ and ignore as many lattice-dependent subtleties as possible. This crude approximation can be trusted provided the only lattice-dependent contributions are order 1 factors (as in the case of $g^+_{sp}$) or factors that do not scale with the rank.

The normalisation factor we need to convert to Weil-Petersson volumes as above is (see Equation \ref{relative_factor})
\begin{equation}\label{eq:norm}
C_{\rm WP}= \left(1 \over \sqrt{2} \right)^{2n}\frac{\sigma(2 + n)}{\sigma(2)\sigma(n)}
\end{equation}
where $\sigma(D) \equiv 2^{(D + 1)/2}\prod_{j=1}^{D-1}\left({(2 \pi)^{{j + 1 \over 2}} \over \Gamma({j + 1 \over 2})} \right)$. Recall that the first factor of ${1 \over \sqrt{2}}^{2n}$ is the conversion factor to the Weil-Petersson metric worked out in Appendix \ref{app:norm}, and the second factor $\frac{\sigma(2 + n)}{\sigma(2)\sigma(n)}$ is the conversion factor derived to go from the Hirzebruch-Mumford volume ${\rm vol}_{HM}$ to the ``Lie algebraic'' volume ${\rm vol}^{tr}$, as in \cite{Moore}.

Now let us examine the behavior of some volumes. If for now we completely ignore the (important-but-lattice-dependent) factors $\frac{2}{g^+_{sp}} |{\rm det} L|^{(n+2 + 1)/2}\prod_p \alpha_p(L)^{-1}$, we can look at the behavior with $n$ of $\beta \equiv C_{\rm WP}\prod_{k=1}^{n+2}\pi^{-k/2}\Gamma(k/2)$ numerically. At $n=1$ we simply have $\beta = {1 \over \pi}$ and it decreases until $n=17$ where it reaches \begin{equation}\beta = \frac{24329988412181570252900390625}{1048576 \pi ^{73}} \simeq 10^{-14},\end{equation} after which it starts increasing dramatically. (As it goes from $n=27$ to $n= 28$, it crosses over from $\beta <1$ to $\beta >1$). See Figure 1.
%\begin{figure}
%\includegraphics[scale=0.45]{betaplot.pdf}
%\caption{The red curve is $\log(\beta)$ versus $n$. The other curves depict $\log({\rm vol}_{\rm WP}(O(L)))$ versus $n$ for the lattices $L= {\rm II}_{2, 2 + 8s}$ (orange), $T_{2, 2 + 8s}$ (green), $F^2_{2, 3 + 8s}$(blue), $F^7_{2, 3 + 8s}$(purple).}
%\end{figure}

%\begin{figure}
%\includegraphics[scale=0.45]{betaplot_geom.pdf}
%\caption{Here we reproduce the plot in Figure 1, with each volume is multiplied by an additional factor of $n!/\pi^n$. This is the analogue of the multiplicative factor $h_{2, 1}!/\pi^{h_{2, 1}}$ that is relevant for the index density (cf. \eq{factor}).
%required to relate the geometric factor in the index density to the Weil-Petersson volume
%}
%\end{figure}

\begin{figure}
  \centering
  \begin{tabular}{cc}
  \includegraphics[width=.45\textwidth]{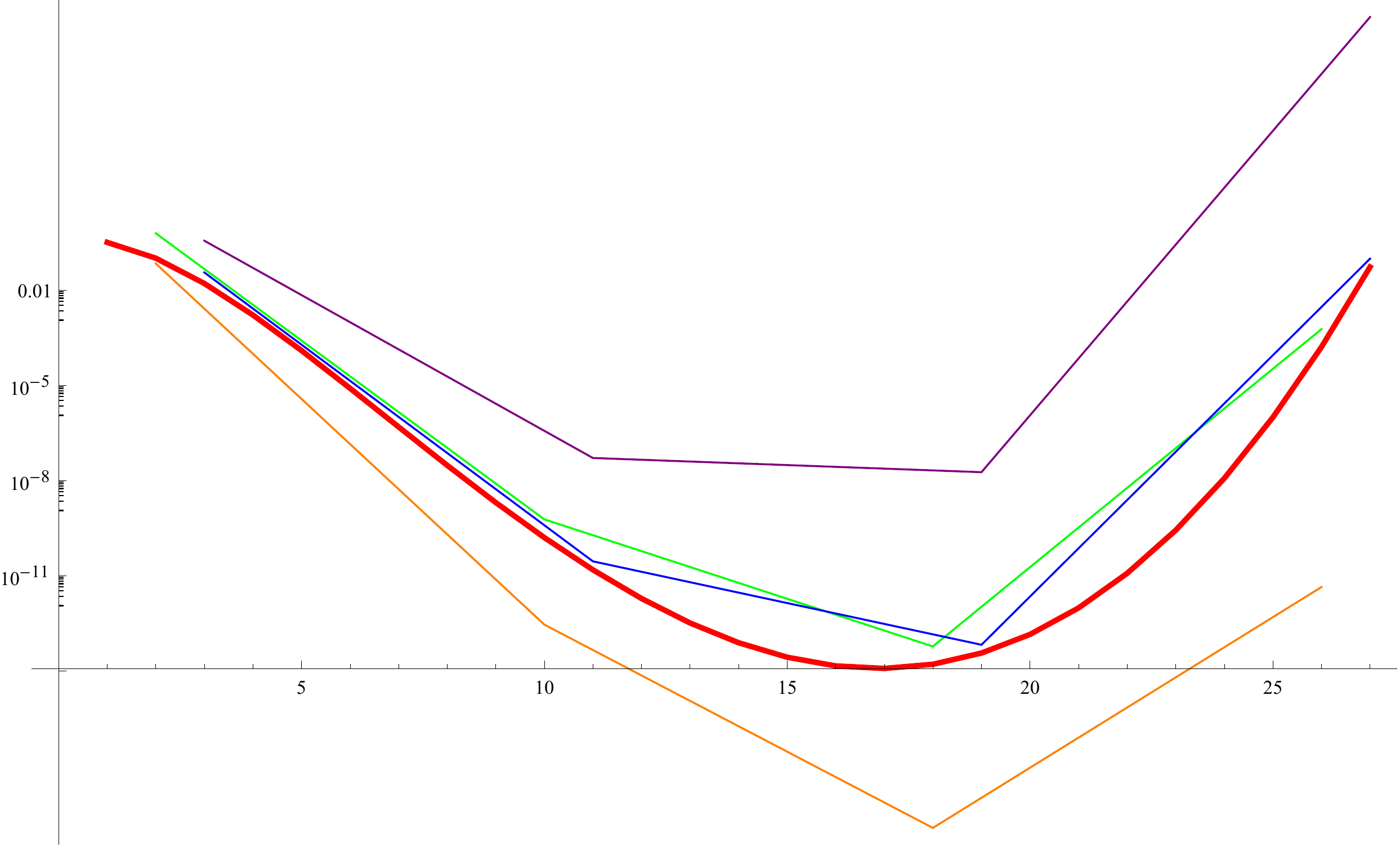}\hspace{5pt}&
  \includegraphics[width=.45\textwidth]{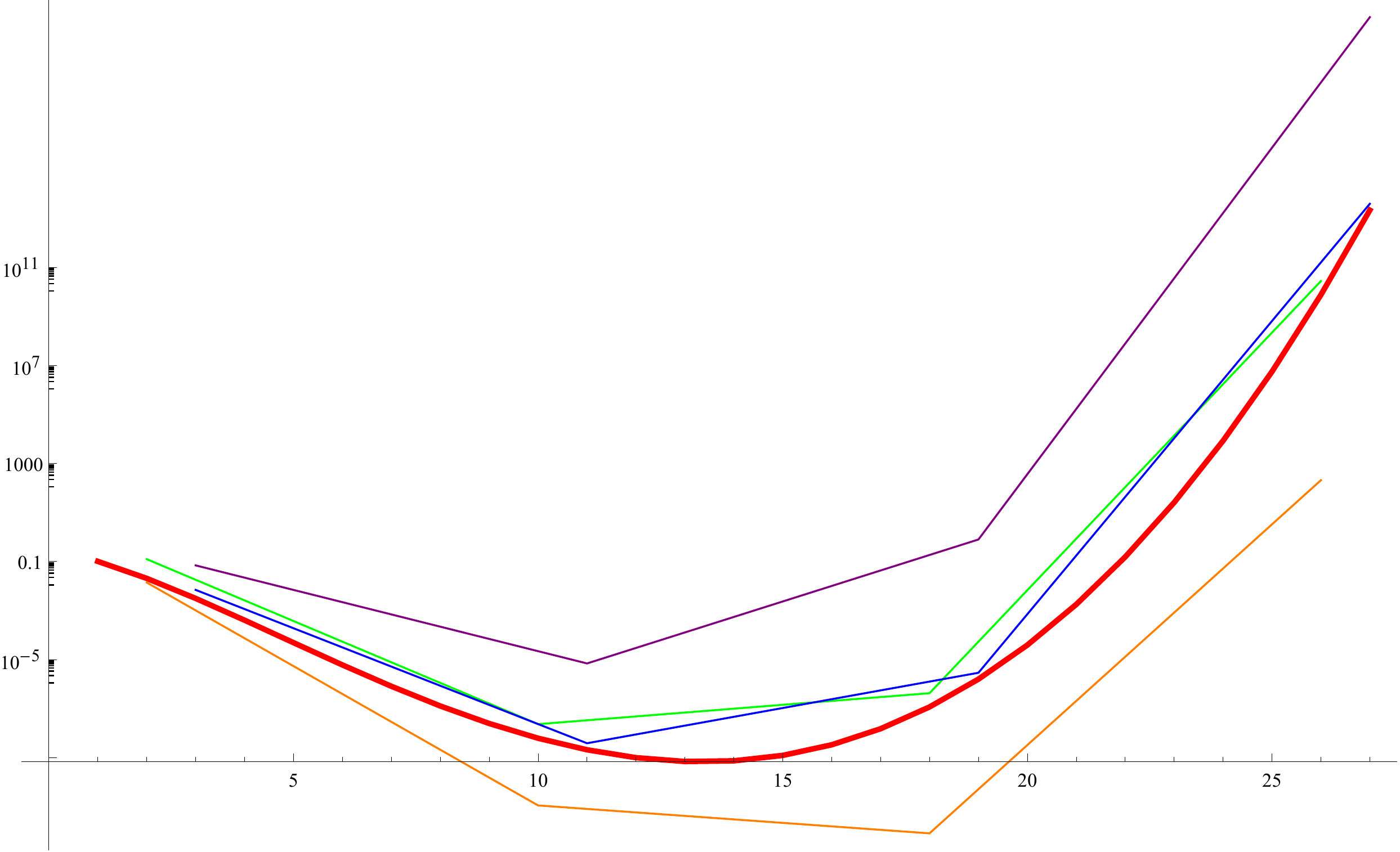}
  \end{tabular}
  \caption{\label{fig:n}In the left figure, the red curve  plots $\log(\beta)$ versus $n$, and the other curves depict $\log({\rm vol}_{\rm WP}(O(L)))$ versus $n$ for the lattices $L= {\rm II}_{2, 2 + 8s}$ (orange), $T_{2, 2 + 8s}$ (green), $F^2_{2, 3 + 8s}$(blue), $F^7_{2, 3 + 8s}$(purple). In the right figure each volume is multiplied by an additional factor of $n!/\pi^n$. This is the analogue of the multiplicative factor $h_{2, 1}!/\pi^{h_{2, 1}}$ that is relevant for the index density (cf. \eq{factor}).  }
  \end{figure}

Next we turn to the behavior with $n$ of the  lattice-dependent factors: $\frac{2}{g^+_{sp}} |{\rm det} L|^{(n+2 + 1)/2}\prod_p \alpha_p(L)^{-1}$. For illustration, Figure \ref{fig:n} displays the differences between $\beta$ and ${\rm vol}_{\rm WP}(O(L))$ for some even unimodular lattices of signature $(2, 2 + 8b)$. Recall that the latter are the lattices of the form $L={\rm II}_{2, 2 + 8b} = 2 \Gamma^{1, 1} \oplus b E_8(-1)$ (i.e. two copies of the hyperbolic lattice and b copies of the $E_8$ root lattice), and the volumes for these have been computed in \cite{GHS, Moore}. In terms of the Bernoulli numbers $B_n$ and the function $(2n)!! \equiv 2.4. \ldots 2n$ they are given by ${\rm vol}_{HM}(O({\rm II}_{2, 2 + 8b})) = 2^{-(4 b + 1)}{B_2 \ldots B_{8b + 2} \over (8 b + 2)!!}{B_{4 b + 2} \over 4 b + 2}$.

\begin{table}[h!]\begin{center}
\begin{tabular}{ l | r r r r r }\toprule
  $b$ & 0 & 1 & 2 & 3 & 4 \\\midrule
  $\beta$ & 0.101 & $1.5\times 10^{-10}$ & $1.6\times 10^{-14}$ & $1.6\times 10^{-4}$ & $3.6\times 10^{25}$ \\\midrule
  ${\rm vol}_{\rm WP}(O(L))$ &  0.068 & $2.8\times 10^{-13}$ & $1.1\times 10^{-19}$ & $4.3 \times 10^{-12}$ & $3.8\times 10^{15}$ \\\bottomrule
\end{tabular}\end{center}
\end{table}
We can see the qualitative similarities between the behavior with $n$ of $\beta$ and the full ${\rm vol}_{\rm WP}$ for this class of lattices; in fact, the scaling with $n$ is exacerbated by the inclusion of the local factors. %It is tempting to assume that such behavior will persist regardless of the details of the underlying lattices.
To get a sense of whether this is a generic feature, let us consider the behavior of some of the other examples computed in \cite{GHS}. See Figure \ref{fig:n} for plots of their volumes.

\textbf{The Lattices $L=T_{2, 2 + 8 b}$: }

Consider now the lattices of signature $(2, 2 + 8b)$ that are of the form $T_{2, 2 + 8 b} \equiv \Gamma^{1, 1} \oplus \Gamma^{1, 1}(2) \oplus b E_8(-1)$. Notice that one of the hyperbolic lattices has been rescaled relative to the other.
It turns out that these volumes are those of the even unimodular examples, multiplied by an additional factor $(2^{4b + 1} + 1)(2^{4 b + 2}-1)$, following logic similar to that which we employed in Section \ref{sec:FHSV}:
\begin{table}[h!]\begin{center}
\begin{tabular}{ l | r r r r r }\toprule
  $b$ & 0 & 1 & 2 & 3 & 4 \\\midrule
  ${\rm vol}_{\rm WP}(O(L))$ &  0.617 & $5.9\times 10^{-10}$ & $5.8\times 10^{-14}$ & $5.8 \times 10^{-4}$ & $1.3\times 10^{26}$ \\\bottomrule
\end{tabular} \end{center}
\end{table}

\textbf{The Lattices $L=F^{d}_{2, 3 + 8 b}$: }

As a final example, consider the set of lattices of signature $(2, 8 b + 3)$ of the form $F^{d}_{2, 3 + 8 b} = 2 \Gamma^{1, 1} \oplus b E_8(-1) \oplus \langle -2 d \rangle$. When $b=2$, this moduli space is (almost) that of polarized K3 surfaces of degree $2d$ \cite{GHS}. For $d>1$\footnote{The $d=1$ case is identical up to an additional factor of 2.} the answer is
\begin{equation}
{\rm vol}_{\rm HM}(O(L)) = \left({d \over 2}\right)^{{8 b + 4 \over 2}} \prod_{p|d}(1 + p^{-{8b + 4 \over 2}}){|B_2 \ldots B_{8 b + 4}| \over (8b + 4)!!}.
\end{equation}

The volumes after normalising are given for low-lying $b$ and several values of $d$ below:

\begin{table}[h!]\begin{center}
\begin{tabular}{ l | r r r r r }\toprule
  $b$ & 0 & 1 & 2 & 3 & 4 \\\midrule
  ${\rm vol}_{\rm WP}(d=2)$ &  0.036 & $2.8\times 10^{-11}$ & $6.6\times 10^{-14}$ & 0.097 & $1.1\times 10^{31}$ \\\midrule
  ${\rm vol}_{\rm WP}(d=7)$ &  0.359 & $5.1\times 10^{-8}$ & $1.8\times 10^{-8}$ & $4.0 \times 10^{6}$ & $7.0\times 10^{40}$ \\\midrule
  ${\rm vol}_{\rm WP}(d=100)$ &  93 & 0.43 & 6440 & $5.9\times 10^{22}$ & $4.3 \times 10^{61}$ \\\bottomrule
\end{tabular}\end{center}
\end{table}

Notice that if $d$ is large and has many prime factors, the volumes start out fairly large, in contrast to the other examples.
In general, any contribution from the lattice-dependent factors that scales with the rank (e.g. the factor $|{\rm det} L|^{(3 + n)/2}$ for lattices that are not unimodular) can have a pronounced quantitative effect. \\

 From the above numbers, plotted in Figure \ref{fig:n}, we can see the qualitative similarities between the behavior with $n$ of $\beta$ and the full ${\rm vol}_{\rm WP}$ for these lattices. This provides some evidence that the behaviours of $\beta$ with $n$ is indicative of how the volumes depend on the lattice ranks. Given the decrease in $\beta$ in the range $n < n^*$, below some minimum $n^*$, it is tempting to conclude that moduli space volumes (at least for lattices of this signature) are tiny for the ranks of relevance to string theory. More precisely, the volumes reach their minimum value when the ranks of the lattices approximately coincide with known ranks of complex structure moduli spaces in threefold compactifications with non-generic holonomy, based on quotients of $K3 \times T^2$ or $T^6$. In particular, as mentioned above, lattices of these signatures govern the moduli space of certain polarized K3 surfaces.

 In other words, in the case when the Calabi-Yau moduli space is of this form (of symmetric spaces), our computation suggests an increase of the importance of the damping effect coming from the geometric factor as the relevant Hodge number (governing the dimension of the flux space) increases.
 Extrapolating this effect to more general Calabi-Yaus, it suggests that the previously ignored geometric factor could be a cause for caution
 when drawing the conclusion that the landscape is dominated by Calabi-Yaus with extremely large Hodge numbers. See for instance \cite{Taylor:2015xtz}.
This said, our analysis does not provide strong evidence that the possible damping effect of the geometric factor is generically more dominant than the growth effect from the flux counting factor as the Hodge number increases.

\subsection{Self-Mirror Calabi-Yaus}
The special form of the FHSV moduli space makes it an ideal intermediate case for studying moduli space geometry between the compactifications preserving $\mathcal{N}=4$ supersymmetry  and compactifications on more generic (simply connected) Calabi-Yaus with $SU(3)$ holonomy. Recall that the self-mirror property of the Enriques Calabi-Yau makes it possible to perform perturbatively exact computation on the moduli space geometry.
In the spirit of exploring simpler $\mathcal{N}=2$ compactifications, it is therefore amusing to note the existence of other self-mirror Calabi-Yaus. This property implies that, just as for the FHSV model, their moduli space geometries are protected from certain quantum corrections and could therefore furnish examples of Shimura varieties which are amenable to exact volume computations. %\textcolor{red}{to specify: in which sense is that implied? } \textcolor{blue}{NP: I edited the preceding sentence, see if you agree. Maybe something stronger is true, however. E.g. there may be an argument like self-mirror $\rightarrow$ no instanton corrections $\rightarrow$ constant curvature moduli space $\rightarrow$ moduli space is symmetric space. The Schoen CY would be another such example that people study a lot... Either way, may not be worth dwelling on.}

Two interesting and natural classes of self-mirror Calabi-Yaus have been recently studied in \cite{HK1, HK2} following the work of \cite{OS}. These are the 14 Calabi-Yaus \footnote{up to deformation equivalence} (8 of the so-called type $K$ \cite{HK1, HK2} and 6 of type $A$\cite{OS}) with infinite fundamental group and so enjoy holonomy further reduced from $SU(3)$. The former are realized as  free quotients of $K3 \times T^2$ while the latter are realized as quotients of abelian threefolds. Of course, the FHSV model is the most well-studied representative of the type $K$ varieties.  Focusing on the three-folds of type $K$, we expect that the type $K$ moduli spaces in most cases will be of the form
\be
 \mathbb{H}/\Gamma_E \times \left(O(\Gamma^{2, n})\backslash O(2, n)/(O(2)\times O(n)) \right)
\ee
In the above, $\Gamma_E$ denotes an appropriate congruence subgroup of $SL(2, \mathbb{Z)}$ depending on the quotient group. If we denote the Calabi-Yau by $(S \times E)/G$ where $S$ is a K3 surface and $E$ an elliptic curve, we have $n = \text{rank}H^2(S, \mathbb{Z})^G$ and $\Gamma^{2, n} \simeq H^2(S, \mathbb{Z})^G$, the $G$-invariant part of the integral K3 cohomology lattice. See Section 3.4 of \cite{HK1} for details.

The upshot is that, given the detailed description of these manifolds given in \cite{HK1, HK2}, one can re-do the volume computation of the previous section by computing the appropriate local factors associated to the (even but not unimodular) lattices $H^2(S, \mathbb{Z})^G$; we expect such a computation would yield numerics comparable to those of the previous subsection.

\section*{Acknowledgements}

We thank T. Grimm, S. Kachru, D. Morrison and W. Taylor for useful conversations. The work of M.C. is supported by ERC starting grant H2020 \#640159 and NWO vidi grant (number 016.Vidi.189.182). The work of N.M.P is supported by a Sherman Fairchild Postdoctoral Fellowship. This material is based upon work supported by the U.S. Department of Energy, Office of Science, Office of High Energy Physics, under Award Number DE-SC0011632. We thank the Isaac Newton Institute for hospitality during the conception of this project. Part of this work was  performed at the Aspen Center for Physics, which is supported by National Science Foundation grant PHY-1607611. N.M.P also thanks Perimeter Institute for hospitality while this work was being completed. Research at Perimeter Institute is supported by the Government of Canada through the Department of Innovation, Science and Economic Development and by the Province of Ontario through the Ministry of Research and Innovation.

\appendix
\section{The index density and the Weil-Petersson metric}\label{app:derivation}
In this appendix, we will briefly review the derivation of the index density in \cite{AshokDouglas} following the method of \cite{Denef} to emphasize the appearance of the (canonically normalised) Weil-Petersson metric in the final formula. In Appendix \ref{app:norm} we will subsequently compare the canonical normalisations of the Weil-Petersson and Zamolodchikov metrics to fix the numerical constants appearing in the volume formula. We begin by recalling the Weil-Petersson metric for a Calabi-Yau $n$-fold $X$.
Denote by $\mathcal{H} \rightarrow \mathcal{M}$ the first Hodge bundle  over the Calabi-Yau complex structure moduli space.
It has  fiber $H^n(X)$ of complex dimension $2(h_{n-1, 1}+ 1)$. The $n$-form $\Omega$ is a local, nonzero holomorphic section of the projectivization of $\mathcal{H}$.
%textcolor{blue}{In these conventions I think the 'i' in the subsequent equation is not really there, right? If you agree, please remove the i at will from here and from its counterpart in Appendix C}\MC{I disagree since $\int_M \Omega\wedge \bar \Omega$ is purely imaginary.}
%\MC{added here:}
The K\"ahler potential is
\be
{\cal K} =-\log \left(i \int_M \Omega\wedge \bar \Omega \right)
\ee%}
and the corresponding Hermitian metric is given by
\be\label{def:WP}
G_{{\rm WP},A\bar B} =- {\pa^2 \over \pa_{A} \pa_{{\bar B}}} {\rm log}\left(i \int_M \Omega\wedge \bar \Omega \right) .
\ee

To account for the overall scaling ambiguity of $\Omega$, we introduce a line bundle
${\cal L}$ with  metric $e^{\mathcal{K}}$ and first Chern class\footnote{Here, we follow the conventions of \cite{AshokDouglas}. It is also common to include the factor of ${1 \over \pi}$ directly into the normalisation of $\omega^{\rm WP}$, as in \cite{LuSun}. We will leave this  factor explicit. In particular, it will reemerge in the index density as $\sim {\rm det}(\omega/\pi)$.}
\be
{\omega^{\rm WP}\over \pi}:= c_1(\mathcal{H}) = {i \over 2 \pi}\partial \bar{\partial}{\cal K}.%=: {1 \over 2 \pi}\partial \bar{\partial}{\rm log}\mathcal{K}(z, \bar{z})
\ee
One can thus
  view $\Omega$ as a local, nonzero holomorphic section of $\mathcal{H}\otimes\mathcal{L}$.%, where $\mathcal{L}$ is the line bundle whose first Chern class is the usual K{\"a}hler class and whose associated Hermitian metric is $e^{\mathcal{K}}$.
%In local complex coordinates, we can write

%\be
%{\left(\omega^{\rm WP} \over \pi\right)}^h = {\left(i \over 2 \pi \right)}^h h! ({\rm det} G_{{\rm WP}, A\bar{B}})\,dz_1 \wedge d\bar{z_1} \ldots dz_{h}\wedge d\bar{z_h}
%\ee where $h :=  \text{dim}_{\mathbb{C}}(\mathcal{M}) $.

%\bigskip
The index density arises from a parallel with the following simple expression for the number of zeros of a function $f(x)$ in one real variable $x$:
\begin{equation}
\#\left\lbrace x|f(x)=0 \right\rbrace = \int dx \ \delta(f(x))|f'(x)|.
\end{equation}
Analogously, the number of flux vacua in the complex structure moduli space of an Calabi-Yau fourfold $Z$ is given by
\begin{equation}
N_{\rm vac} = \sum_N \int_{\mathcal{M}} d^{2h}z \ \delta^{2h}(DW_N)\,|{\rm det} D^2 W_N|,
\end{equation}
where $h :=  \text{dim}_{\mathbb{C}}(\mathcal{M})$, $W_N$ is the flux superpotential determined by $N$, and the sum is taken over fluxes satisfying (\ref{eq:fluxes}).
The index density is then given by the following approximate quantity:
\begin{equation}
I_{\rm vac} = \int d^b N \int_{\mathcal{M}} d^{2h}z \ \delta^{2h}(DW_N)\,{\rm det} D^2 W_N,
\end{equation}which is a good approximation in the large flux limit (see \cite{Denef:2004ze} for a discussion about subleading corrections).

In the F-theory context, we have $D_\mu W_{N,\mu} = N^I \Pi_{I \mu}$ where the periods in a fixed homology basis are, as usual, $\Pi_I = \int \Sigma_I \wedge \Omega_4$ and $\Pi_{I \mu}:= e^{\mathcal{K}/2}(\partial_{\mu} + \partial_{\mu} \mathcal{K})\Pi_I(z)= e^{\mathcal{K}/2}D_{\mu}\Pi_I(z)$. Note that here $W$ denotes the rescaled superpotential. See (2.1)-(2.2) of \cite{Denef:2004ze}. 
%The K{\"a}hler potential is related to the usual K{\"a}hler form $\omega$ via $\omega = {i \over 2} \partial \bar{\partial}\mathcal{K}$\MC{still confused about the factor  of 2}.

To evaluate the corresponding index density it is useful to define, as a computational tool, an (a priori) \textit{auxiliary} metric on the moduli space ${\cal G}_{\mu\nu}:= -P_{I\mu} Q^{IJ}P_{J\nu}$ with $Q^{IJ} = Q_{IJ}^{-1}$ the inverse of the intersection matrix on $H^4(Z, \mathbb{Z})$. Here $\mu, \nu$ are to be understood as indices for real coordinates $\mu,\nu = 1 \ldots 2h$. We also have, following the prescription of \cite{Denef}, $P_{I\mu} =\Pi_{I\mu}$ for $\mu = 1, \ldots, h$ and $P_{I\mu} = \bar{\Pi}_{I(\mu-h)}$ for $\mu = h + 1 \ldots 2h$. In complex coordinates one can compute, using Griffiths transversality, that the metric components are ${\cal G}_{AB} = 0 = {\cal G}_{\bar{A}\bar{B}}$ and ${\cal G}_{A\bar{B}} = -e^{\mathcal{K}}\int D_A \Omega \wedge D_{\bar{B}}\bar{\Omega} = \partial_A \partial_{\bar{B}}\mathcal{K}$ and ${\cal G}$ is Hermitian. Note that this is precisely the Weil-Petersson metric \eq{def:WP}, namely ${\cal G}=G$.

A covariant derivative $\nabla$ with respect to this auxiliary metric must satisfy the condition
\begin{equation}\label{eq:covariant}
P_{I\mu}Q^{IJ}\nabla_{\nu}P_{J\rho} = 0.
\end{equation}
We can rewrite this equation using the definition of $P_{I\mu}$ and passing to complex coordinates as
\begin{equation}
e^{\mathcal{K}/2}D_{A}\Pi_I(z)Q^{IJ}\nabla_\mu \left( e^{\mathcal{K}/2}D_{\bar{B}}\bar{\Pi}_I(z)\right) = 0.
\end{equation}
We stress that such a covariant connection is a connection on $T\mathcal{M}\otimes\mathcal{L}$, where $\mathcal{L}$ is the line bundle on $\mathcal{M}$ of which the supergravity potential is a section: in our conventions, $c_1(\mathcal{L}) = [{\omega^{\rm WP} \over \pi}]$, the curvature form of $\mathcal{L}$, and so the Weil-Petersson volume may be expressed as\footnote{In contrast to \cite{AshokDouglas}, we choose conventions to work with dimensionless quantities from the outset. Therefore, the factors of $-1/M_{pl}^2$ required in \cite{AshokDouglas} to render quantities like $c_1(\mathcal{L})$ dimensionless do not appear in our formulas.}
\begin{equation}
{\rm vol}_{\rm WP}(\mathcal{M}) = {(\omega^{\rm WP})^h \over h!} = {\pi^h \over h!}c_1(\mathcal{L})^h.
\end{equation} Expanding Equation (\ref{eq:covariant}) and applying Griffiths transversality, following \cite{Denef}, shows that the covariant connection on $T\mathcal{M}\otimes\mathcal{L}$ is exactly the standard Levi-Civita K{\"a}hler connections, i.e. the auxiliary metric recovers the Weil-Petersson metric on moduli space.

One can then follow the derivation in \cite{Denef}, by constructing the generating function
\begin{equation}
Z(t)= \int d^{4h+4} N \ e^{-t/2 N^I Q_{IJ}N^J}\int_{\mathcal{M}}d^{2h}x \ \delta^{2h}(N^I P_{I\mu})\,\text{det}(\nabla_{\mu}(N^J P_{J\nu}))_{\mu\nu}
\end{equation}
and expressing the index density as $I_{\rm vac}(Q_c) = {1 \over 2\pi i}\int{dt \over t}e^{-t Q_c}Z(t)$ with the contour passing the pole $t=0$ on the left; recall $-{1 \over 2}N^I Q_{IJ}N^J = Q_c$. We sum over the repeated $I, J$ indices, with $I, J = 1, \ldots b = 4h+4$. Notice that the integral over fluxes in $Z(t)$ is now taken over the full $(4h+4)$-dimensional Euclidean space, with the Laplace transform enforcing the bound on fluxes. Rewriting the delta function and determinant factors as integrals over extra Grassmann variables leads to the integral over continuous fluxes to become a Gaussian integral. The series of simplifications outlined in  \cite{Denef} then results in the final expression
\begin{equation}
I_{\rm vac} = {1 \over \sqrt{{\rm det} Q_{IJ}}}{(2 \pi Q_c)^{2h + 2} \over (2h + 2)!}\int_\mathcal{M}e(\nabla)
\end{equation} where $e(\nabla)= \text{Pf}\left(\mathcal{R}_{\underline{\mu}\underline{\nu}}/2\pi\right) = {1 \over \pi^h}\,\text{det}(\mathcal{R} + \omega \cdot \textbf{1})$. The first equality is in terms of the Pfaffian of the curvature form on $\mathcal{M}$ in an orthonormal frame (represented by underlined indices) with respect to $G_{\mu\nu}$. Its appearance follows from identifying the Grassmann integral representation of the Pfaffian directly in the aforementioned manipulations after performing the $\int d^{4h + 4}N$ integral.

Crucially, the auxiliary metric appearing in the derivation coincides \textit{precisely}  with the physical metric on moduli space, including the proper normalisation for the Weil-Petersson metric: $\mathcal{L}$, with metric $e^{\mathcal{K}}$, captures the scaling ambiguity of the top-degree form $\Omega$ and its first Chern class is unambiguously defined\footnote{Recall that the first Chern class map produces a certain, fixed constant multiple of the trace of the curvature operator associated to a chosen connection on $\mathcal{L}$, via Chern-Weil theory.} and gives the metric associated to the K{\"a}hler connection. If we drop the curvature term in the index density, then we find ${(\omega^{\rm WP})^h/\pi^h} = {h! \over \pi^h}\text{vol}_{WP}(\mathcal{M})$ as claimed.

It is frequently stated that the Weil-Petersson and Zamolodchikov metrics coincide. We claim that in fact, these metrics differ in their canonical (constant) normalisations.  In Appendix \ref{app:norm} we quantify this discrepancy and thereby fix the scale of the metric.

\section{Some number theoretic objects}\label{app:padic}
In this appendix, we elaborate on the definition and computation of the remaining ingredients in the lattice volume formula.

First, we recapitulate the definition of proper spinor genus in \cite{Kitaoka}, which is a definition particularly suitable for computations \cite{Fiori}. The reader who wants to learn more about these quantities is also advised to consult \cite{SPLAG, Cassels} for more conceptual definitions and further references.

Recall that the spinor norm $\theta: O(V) \rightarrow F^{\times}/(F^{\times})^2$ where $V$ is a quadratic space over $F$. The map is explicitly given by $\theta(\sigma):= Q(v_1)Q(v_2)\ldots Q(v_n)$ where $\sigma = \tau_{v_1}\tau_{v_2}\ldots \tau_{v_n} \in O(V)$ is written as a product of elementary reflections $\tau_{v_i}$ with respect to basis elements $v_i \in V$, and $Q$ denotes the quadratic form.

The genus $\text{gen}(L)$ of a lattice $L$ on a quadratic space $V$ is the set of lattices $M$ on $V$ such that for some $\sigma_p \in O(V_p)$
\begin{equation}\nonumber
M_p = \sigma_p(L_p) {\text{ for every prime }} p.
\end{equation}
A genus is called proper if one replaces $O(V_p)$ with $O^+(V_p)$, the subgroup of elements that preserve the orientation of all positive-definite planes.
The spinor genus $g_{sp}(L)$ of $L$ is the set of lattices $M$ such that for some $\eta \in O(V)$ and some $\sigma_p \in O'(V_p)$ we have
\begin{equation}\nonumber
\eta(M)_p = \sigma_p(L_p) {\text{ for every prime }} p.
\end{equation}
The group $O'(V_p)$ is the kernel of the spinor norm from $O^+(V_p)$ to $(\mathbb{Q}^{\times}_p)/(\mathbb{Q}^{\times}_p)^2$. Finally, the proper spinor genus uses the same definition except with $\eta \in O^+(V)$.

Cor 6.3.1 of \cite{Kitaoka} establishes that $g^+_{sp}(L)$, which appears in the volume formula, is always a power of 2. Cor 6.3.2 of \cite{Kitaoka} further establishes how the numerical value can be obtained by a finite computation. For our purposes, we note that for a lattice $L$ of rank $(2, n)$ containing at least one copy of the hyperbolic plane as a direct summand, one can prove that ${1 \over g^+_{sp}} = 1$ using results of \cite{Kneser} (see \cite{GHS}).

The definition of a local density $\alpha_p(S)$ of a quadratic form over a number field $F$ given by a matrix $S \in \text{Mat}_{n\times n}(F)$ is

\begin{equation}
\alpha_p(S):= {1 \over 2}\text{lim}_{r \rightarrow \infty}p^{-r n (n-1)/2}\vert \left\lbrace X \in \text{Mat}_{n \times n}(\mathbb{Z}_p) \text{ mod } p^r; X^t S X \equiv S \text{ mod } p^r \right\rbrace \vert.
\end{equation}
As before, $\mathbb{Z}_p$ denotes the $p$-adic integers. In general, such representation densities serve to assign a volume to sets of isometric embeddings $\text{Isom}(L_1, L_2)$ for lattices over a ring $R$.
To compute these local densities, one needs to know the Jordan decomposition of $L$ over $\mathbb{Z}_p$. Let us elaborate on this.

We call a lattice $L$ over a ring $R$ $a$-modular, where $a$ is an invertible fractional ideal of $R$, if $\text{Hom}_R(L, R) = a^{-1}L$ or, equivalently, $\text{Hom}_R(L, a) = L$. Theorem 4.3.5 in \cite{Fiori} tells us that every lattice $L$ over a p-adic ring $R$ may be written as $L = \oplus_i L_i$ where $L_i$ are $a_i$-modular lattices and each $a_i$ is distinct. This is referred to as a Jordan decomposition of $L$. Jordan decompositions are in general not unique, and one must in general have a method to determine all Jordan decompositions of a given lattice to compute the local densities. However, let us say we have two Jordan decompositions of a lattice over a p-adic ring $R$: $L = \oplus_{i=1}^{r_1}L_i = \oplus_{j=1}^{r_2} K_j$, such that $L_i$ are $a_i$-modular and $K_j$ are $b_j$-modular. Let us also suppose that $a_{i_1}|a_{i_2}$ for $i_1 < i_2$ and $b_{j_1}|b_{j_2}$ for $j_1 < j_2$. Then there is a  uniqueness result for Jordan decompositions (see Theorem 4.3.14 of \cite{Fiori} for the precise statement) that in particular tells us $r_1 = r_2, a_i = b_i, \text{rk}(L_i) = \text{rk}(K_i)$ and $L_i \simeq K_i$ if $p \neq 2$.

The algorithm for computing local densities goes as follows \cite{Fiori, Kitaoka}.

\begin{enumerate}
\item Any Jordan decomposition of a lattice expresses the lattice in terms of $a$-modular summands. One can relate the local density of an $a$-modular lattice to that of a unimodular lattice.
\item One can relate the local density of a unimodular lattice to the local densities of certain lattices that have rank at most 4.
\item Computing the local densities for the low-rank lattices in the previous item may be done explicitly.
\item Finally, the computation of local densities for an arbitrary lattice requires enumeration of all Jordan decompositions of the lattice and the computation of the local factors for the corresponding $a$-modular pieces as above. In this work, we work with simple lattices with simple, unique Jordan decompositions and refer to \cite{Fiori} for discussions of the more general case.
\end{enumerate}
Many local densities of interest in this paper have already been computed in \cite{GHS} and \cite{Moore} to which we refer for the precise formulas; we only modify their results slightly using the recipe of \cite{Fiori} when needed.

For example, for every prime $p$, the even unimodular lattices of signature $(2, 2 + 8b)$ over $\mathbb{Z}_p$ are given by direct sums of hyperbolic planes. For another example, consider the lattice $T_{2, 2 + 8b} = U \oplus U(2) \oplus b E_8$. When $p=2$, we have $T_{2, 2 + 8b}\otimes \mathbb{Z}_2 = (4 b + 1)\Gamma^{1, 1} \oplus \Gamma^{1, 1}(2)$ and for $p \neq 2$ we have $T_{2, 2 + 8b}\otimes \mathbb{Z}_2 = (4 b + 2)\Gamma^{1, 1}$ \cite{GHS}.
%The local densities of a lattice may be determined by what is called their Jordan decomposition. Namely, a lattice $\Lambda$ over a $p$-adic ring $R$ can be expressed (in general non-uniquely) as $\Lambda \simeq \oplus_i L_i$ where the $L_i$ are $a_i$-modular lattices for distinct $a_i$. A lattice is called $a$-modular if $Hom_R(\Lambda, R) = \Lambda/a$ ($a=1$ for unimodular lattices). To avoid making the exposition too dense, we will only state some basic facts. Local densities are some prescription to assign volumes to the sets of isometric embeddings between two lattices that preserve the quadratic form. Without worrying too much about the precise definition, we will mention that there is a recipe for computing these local densities \cite{Fiori}. One must have some combinatorial procedure for finding all possible Jordan decompositions of a lattice. There are then ways to compute local densities of $a$-modular lattices using local densities of unimodular lattices (see Proposition 4.4.4 of \cite{Fiori}). The latter problem may be further reduced to the problem of computing local densities of certain lattices with rank at most 4 (see Theorem 4.4.11 of \cite{Fiori}). The densities of these special cases are determined in Theorem 4.4.18 of \cite{Fiori}. The result for an arbitrary lattice by putting together these ingredients may be found in Theorem 4.4.28 of \cite{Fiori}. In this paper, we will be dealing with lattices that are direct sums of simple $a$-modular lattices, for which we can readily compute the results.

\section{Relation Of The Weil-Petersson And Zamolodchikov Metrics}\label{app:norm}

For the calculation in the main text it is crucial that  we carefully compare the
normalisation between the Hirzebruch--Mumford metric and the Weil--Petersson metric that leads to the pre-factor in \eq{relative_factor}.
To do so, we compare the canonical normalisations of the Weil--Petersson and Zamalodchikov metrics, where the latter is
 determined using the arguments of \cite{Moore}.

  \subsection{Zamolodchikov Metric vs Weil--Petersson Metric}

  \def\pa{\partial}
  \def\a{\alpha}
  \def\s{\sigma}
  \def\m{\mu}
  \def\n{\nu}
  \def\RR{\mathbb{R}}
  \def\til{\tilde}

We revisit the original computation of \cite{CHS}, keeping careful track of overall
normalisations. Although we focus on the complex structure moduli space, the derivation is
completely analogous for the complexified K\"ahler moduli.

First consider the WP metric for the moduli space of complex structures of the Calabi-Yau
three-fold $M$. We use  $l,m$ to denote indices for real coordinates $X^l$. Using the
normalisation of \cite{CHS} the natural metric on the space of metrics on $M$ is
\be\label{eq:MetricOnMetric}
\frac{1}{V} \int_M \sqrt{G} G^{ll'} G^{mm'}  \delta G_{lm} \delta G_{l' m'} d^6 x
\ee
Here $V$ is the volume of $M$.
If we choose a complex structure we let  $\mu,\nu$ denote indices of the complex coordinates.
We also denote by $G_{\mu \bar \nu}$ the Hermitian metric such
    that 
\be
ds^2 = {1\over 2}G_{\mu \bar \nu}(dX^\mu \otimes dX^{\bar \nu} + dX^{\bar \nu} \otimes dX^\mu).
\ee
We denote the inverse to $G_{\mu\bar\nu}$ as
\be\label{eq:InvGmmbar}
\begin{split}
G_{\mu \bar \nu} G^{\mu \bar \rho} & = \delta_{\bar \nu}^{\bar \rho}\\
G_{\mu \bar \nu} G^{\rho \bar \nu} & = \delta_{\mu}^{\rho} \\
\end{split}
\ee
Now, the  tangent space to the complex structure moduli space
can be associated with   first order deformations of the metric of the form
 \be\label{eq:MetricDeform}
 ds^2 \rightarrow ds^2 + \left( h_{\mu\nu}(x) dX^\mu \otimes dX^\nu
 + \bar h_{\bar \mu \bar \nu}(x) dX^{\bar \mu} \otimes d X^{\bar \nu}\right) + \mathcal{O}(h^2)
 \ee
 where $ \bar h_{\bar \mu \bar \nu}(x)= (h_{\mu\nu}(x))^*$, and
 the associated Beltrami differential $G^{\mu \bar \rho} \bar h_{\bar \rho \bar \sigma}$
 is harmonic. Evaluating \eqref{eq:MetricOnMetric} on such deformations gives the
 metric:
  \begin{align}\label{eq:CandelasMet}
  G_{\rm WP}(h^1, \bar h^2) =
   {4\over V}\int_M  d^6x \sqrt{G(x)}h^1_{\m\n}(x)\bar{h}^2_{\bar \rho\bar\sigma}(x) G^{\n\bar\sigma}(x)G^{\m\bar\rho}(x).%\\
    \end{align}
    and the other components vanish because it is of type $(1,1)$ on the complex structure moduli space.
    According to \cite{CHS} \eqref{eq:MetricOnMetric} is precisely the normalisation that gives the canonically
    normalised Weil-Peterson form which explains the subscript {WP}.

Now we turn to the Zamolodchikov metric.
As in \cite{Moore}, we denote a CFT $\mathcal{C}$ as a point in the moduli space $\mathcal{M}$ of CFTs
 and study the map from the space $V^{1,1}$ of exactly marginal operators of ${\cal C}$ to the tangent space
 to ${\cal M}$ at ${\cal C}$: $\Psi: V^{1, 1} \rightarrow T_{\mathcal{C}}{\cal M}$. If our CFT's are defined
 by an action (as is the case here) then a path in ${\cal M}$ is determined by a path of actions
 $S[t]$. If
 \footnote{ We denote  the real worldsheet coordinates by  $\sigma^1, \sigma^2$ and the corresponding
    derivatives by $\partial_{1, 2}$, and also define the complex worldsheet coordinates
    $z := \sigma^1 + i \sigma^2, \bar{z}:= \sigma^1 - i \sigma^2$ with $\partial = {1 \over 2}(\partial_1 -i \partial_2)$, etc.
    In particular $d^2z := \frac{i}{2} dz \wedge d\bar z = d\sigma^1 \wedge d \sigma^2$. }
${d \over dt}|_{t=0}S[t] = \int {\cal O}d^2z $  then
 $\Psi({\cal O}) = {\partial \over \partial t}|_{t=0}$ is the tangent vector to the path in moduli
 space. If a tangent vector $v=\Psi({\cal O})$ to $\mathcal{M}$ corresponds
 to the exactly marginal operator  $\cal{O}$ then we define the Zamolodchikov metric by:
\be\label{eq:ZamDef}
\langle \mathcal{O}(z_1) \mathcal{O}(z_2) \rangle := \frac{g_{\rm Z}(v,v) }{  \vert z_1 - z_2 \vert^4 }
\ee
where the LHS is the correlation function on the complex plane $\mathbb{C}$ with the unique $SL(2,\mathbb{C})$
invariant vacuum at $z=0,\infty$.

Now specialize to a supersymmetric non-linear sigma model on a Calabi-Yau threefold $M$. For simplicity we consider background with
    vanishing $B$-field.  Then the bosonic part of the action reads:
    \begin{gather}\begin{split}
    S_0 &= {1\over 2\ell^2} \int G_{lm}(X)  (\pa_1 X^l \pa_1  X^m +\pa_2 X^l \pa_2  X^m ) \, d\s^1\wedge  d\s^2\\
    &= {1\over {2}\ell^2} \int G_{\m\bar\n}(X)  (\pa_1 X^\mu \pa_1  X^{\bar \nu} +\pa_2 X^\mu \pa_2  X^{\bar \nu} ) \, d\s^1\wedge  d\s^2\\
    &= {1\over  \ell^2} \int  G_{\m\bar\n}(X)  (\pa X^\mu \bar\pa   X^{\bar \nu} +\bar \pa X^\mu \pa  X^{\bar \nu} )\,  d\s^1\wedge  d\s^2
\end{split}    \end{gather} where $\ell^2 \equiv 2 \pi \alpha'$.
Because we are considering the metric to leading order in $\alpha'$ it suffices to consider only the bosonic part of the action. The reason for this is that one can check that the contributions of the fermionic terms to the exactly marginal operator lead to terms in the expression for the Zamolodchikov metric that are all higher order in $\alpha'$. One way to prove this is to show that all the fermionic contributions involve integrals over $M$ with extra insertions of curvature tensors and/or covariant derivatives acting on $h^1$ and/or $\bar h^2$. The fermionic terms certainly would need to be taken into account if one computed the $\alpha'$ corrections to the Weil-Peterson metric.

Metric deformations of the form \eqref{eq:MetricDeform} will preserve the CY property and the corresponding
deformation of the action is associated with the  exactly  marginal operator:
  \begin{gather}\begin{split}
   {\cal O}(h)& := \frac{1}{2\ell^2}  h_{\m\n}(X(\s))  (\pa_1 X^\mu \pa_1  X^{ \nu} +\pa_2 X^\mu \pa_2  X^{ \nu} )+\cdots  \\
    &=\frac{2}{\ell^2}  h_{\m\n}(X(\s)) \pa X^\mu \bar \pa  X^{ \nu}+\cdots  \,
   \end{split}    \end{gather}
with a similar formula for ${\cal O}(\bar h)$. Here $+\cdots$ indicates the fermionic contributions.

We can now compare the Zamolodchikov metric \eqref{eq:ZamDef} with the Weil-Peterson metric, at least in the leading order in
the $\a' \rightarrow 0$ limit.  Note that this is sufficient for us since the metric for the Enriques Calabi-Yau does not receive $\a'$-corrections.

 We write
    \be\label{alpha_expand}
    X^\m(\s) =  x^{\mu} + \tilde X^\m(\s);  \;X^{\bar\m}(\s) =  x^{\bar\m} + \tilde X^{\bar\m}(\s).
    \ee
    and subsequently
    \be \nonumber
    G_{\m\bar\n}(X) = G_{\m\bar\n}(x) + O( \tilde X).
    \ee
    The bosonic part of the action in this limit for the sigma model with vanishing B-field is
\be \til     S_0 = G_{\m\bar\n}(x)  {1\over 2\ell^2} \int  (\pa_1 \til X^\mu \pa_1  \til X^{\bar \nu}
 +\pa_2 \til X^\mu \pa_2  \til X^{\bar \nu} ) \, d\s_1\wedge  d\s_2.\\
\ee
    In this free field limit,  we have
    \footnote{The essential fact is that, on the Euclidean plane $(\pa_x^2 + \pa_y^2) \log\vert z\vert^2 = 4\pi \delta^{(2)}(0)$.
    Adding source terms to the action $\int (j_\mu X^\mu + j_{\bar \mu} X^{\bar \mu}) d\sigma^1 \wedge d \sigma^2$ we cancel them
    by shifting
    $$X^\mu(1) \to X^{\mu}(1) + G^{\mu\bar\rho} \int \frac{\ell^2}{2\pi} \log\vert z_1 - z_2\vert^2 j_{\bar \rho}(2) d^2 \sigma. $$  }
   \begin{gather}\begin{split}
     \langle  \til X^{\m}(\s_1,\s_2) \til X^{\bar \n}(\s'_1,\s'_2) \rangle_{\tilde S_0}
  &= \frac{ \int [D\til X] \til X^{\m}(\s_1,\s_2) \til X^{\bar \n}(\s'_1,\s'_2) e^{-\til S_0}}
  {\int [D\til X]   e^{-\til S_0}}  \\
  &=-\frac{\ell^2}{\pi}
   G^{\m\bar\n}(x) \log((\s_1-\s_1')^2+(\s_2-\s_2')^2 ) + O(\alpha') \\
    &=-\frac{\ell^2}{\pi} G^{\m\bar\n}(x)\log|z-z'|^2 + O(\alpha').
    \end{split}\end{gather}
where we recall \eqref{eq:InvGmmbar}. Moreover
    \be \nonumber
    \langle  \til X^{\m}(\s_1,\s_2) \til X^{ \n}(\s'_1,\s'_2) \rangle_{\tilde S_0} = \langle  \til X^{\bar \m}(\s_1,\s_2) \til X^{\bar \n}(\s'_1,\s'_2) \rangle_{\tilde S_0} = O(\alpha').
    \ee
    Hence
    \begin{gather}\begin{split}
   \langle \pa \til X^{\m}(\s_1,\s_2)\pa \til X^{\bar \n}(\s'_1,\s'_2) \rangle_{\tilde S_0}& =-\frac{\ell^2}{\pi}
   G^{\m\bar\n}(x) {1\over (z-z')^2}  + O(\alpha')  \\
   \langle \bar\pa \til X^{\m}(\s_1,\s_2)\bar \pa \til X^{\bar \n}(\s'_1,\s'_2) \rangle_{\tilde S_0} &=
   -\frac{\ell^2}{\pi} G^{\m\bar\n}(x) {1\over (\bar z-\bar z')^2}  + O(\alpha')
    \end{split}\end{gather}
    and
    \begin{align} \langle \pa \til X^{\m}(\s_1,\s_2)\bar \pa \til X^{\bar \n}(\s'_1,\s'_2) \rangle_{\tilde S_0} = \langle \bar \pa \til X^{\m}(\s_1,\s_2) \pa\til X^{\bar \n}(\s'_1,\s'_2) \rangle_{\tilde S_0}= O(\alpha').
        \end{align}

   % \textcolor{red}{Note that this is quite different from eqn (18) of Candelas--H\"ubsch--Schimmrigk.}

    From this we conclude
    \be
    \langle {\cal O}(h^1)(\s) {\cal O}(h^2)(\s') \rangle= \langle {\cal O}(\bar h^1)(\s) {\cal O}(\bar h^2)(\s') \rangle  = O(\alpha') . \nonumber
   % \langle {\cal O}_{\m\n}(\s) {\cal O}_{\bar\m\bar\n}(\s') \rangle = +O(\alpha').
    \ee

    Using again the expansion
    %\eq{alpha_expand}
     and writing
     \be
     \int [{\cal D}X]....  = \int d^6x \sqrt{G(x)} \int   [{\cal D}\til X]....
     \ee
(cf. (17) of \cite{CHS}), we obtain
\be
\begin{split}
\langle \mathcal{O}(h^1)(z_1, \bar z_1) \mathcal{O}(\bar h^2)(z_2,\bar z_2) \rangle &
:=  \frac{ \int [{\cal D}X] \mathcal{O}(h^1)(z_1, \bar z_1) \mathcal{O}(\bar h^2)(z_2,\bar z_2) e^{-S} }
{ \int [{\cal D}X]  e^{-S} } \\
 =\frac{1}{V} \int_M d^6x \sqrt{G(x)} \left(\frac{2}{\ell^2}  \right)^2
 &
  h^1_{\mu\nu}(x) \bar h^2_{\bar\rho \bar \sigma}(x)
 \langle \pa \tilde X^\mu\bar\pa \tilde X^{\nu}(z_1,\bar z_1)
  \bar\pa \tilde X^{\bar \rho}  \pa \tilde X^{\bar \sigma}(z_2,\bar z_2)\rangle_{\tilde S_0} \\
 & = \frac{4}{ \pi^2} \frac{1}{\vert z_1 - z_2 \vert^4}
 \frac{1}{V}\int_M d^6x \sqrt{G(x)} G^{\mu\bar\rho}(x) G^{\nu\bar \sigma}(x)  h^1_{\mu\nu}(x) \bar h^2_{\bar\rho \bar \sigma}(x)  \\
\end{split}
\ee
%

  %  \begin{align}\nonumber
  %    \langle {\cal O}(h^1)(\s) {\cal O}(\bar h^2)(\s') \rangle   & = \int [{\cal D}X]  {\cal O}_{A}(\s) {\cal O}_{\bar B}(\s') e^{-S_0}   \\  \nonumber
  %    & = {1 \over  2 \pi^2} {1\over |z-z'|^4} {1\over V}\int  d^6x \sqrt{G(x)} g_{A,\m\n}(x)g_{\bar A,\bar \rho\bar\sigma}(x) G^{\n\bar\sigma}(x)G^{\m\bar\rho}(x).
  %  \end{align}
  %
  Using the definition \eqref{eq:ZamDef} and comparing
  with \eq{eq:CandelasMet}
    we conclude that the Zamolodchikov metric in the leading order of $\a'$ is simply given by
  \be
   ds_Z^2  = {1\over \pi^2} ds_{\rm WP}^2.
  \ee

\subsection{Consistency check for square tori}

The Zamolodchikov metric for a periodic scalar of radius $R$
 was computed in \cite{Moore} to be $\frac{1}{\pi^2}(\frac{d R}{R})^2$ and the metric for a $2d$-dimensional square torus was similarly given to be $\frac{1}{\pi^2}\sum_{i=1}^{2d}(\frac{d R_i}{R_i})^2$.
 As a test of our proposed normalisation, we will consider the moduli space of a complex abelian variety and consider the pullback of the metric to the sublocus of products of square tori with zero $B$-field.
%\begin{equation}\label{eq:squaretori}
%\sum_i \left(\frac{dR_i}{R_i}\right)^2 = ds^2_{\rm WP}|_{\text{sq \ tori}}
%\end{equation}
The K{\"a}hler deformations of the metric are \cite{CHS}
\begin{equation}
G_{A \bar{B}} \delta w^{A} \delta w^{\bar{B}} = \frac{1}{V}\int_{\mathcal{M}} d^6 z\,  \sqrt{G}\, G^{\mu \bar{\tau}} G^{\rho \bar{\nu}} \left(\delta G_{\mu \bar{\nu}} \delta G_{\rho \bar{\tau}} + \delta B_{\mu \bar{\nu}} \delta B_{\bar{\tau}\rho} \right)
\end{equation} and one may follow the arguments of the previous subsection identically to obtain the same relative normalisation between the metric on the space of complexified K{\" a}hler moduli and the Zamolodchikov metric: 
\begin{equation}
ds^2_Z = {1 \over \pi^2} ds^2_{K}
\end{equation}
so that for a threefold given by a square abelian variety we expect to obtain
\begin{equation}\label{eq:squaretori}
ds^2_K = \sum_{i=1}^{6}\left(\frac{d R_i}{R_i}\right)^2.
\end{equation}
The components of the canonically normalised K{\"a}hler metric are defined for threefolds as \cite{CHS}
\begin{equation}\label{eq:wpmetric}
G_{A \bar{B}} = -\frac{\partial^2}{\partial w^{A} \partial w^{\bar{B}}} \log \left( \int_{\mathcal{M}} J \wedge J \wedge J \right).
\end{equation}

%The K{\"a}hler form is just $J= i \sum_{k, j=1}^d g_{k, \bar{j}}(z, \bar{z}) d z^k \wedge d\bar{z}^j$ such that
%\begin{equation*}
%J^d/d! = 2^d \textrm{ det}(g_{i \bar{j}}) dx^1 \wedge \ldots \wedge dx^{2d}
%\end{equation*} is the usual volume form. Note that in this
%notation $2^d \textrm{ det}(g_{i \bar{j}}) = \sqrt{\textrm{ det}(g_{\mu \nu} )}$. 
We restrict our manifold $T^6$ to be the product $T^2 \times T^2\times T^2$ and plug the factorised K{\"a}hler form into \ref{eq:wpmetric}. Restricting to the locus of square tori with zero B-field, each $T^2$ factor has K\"ahler modulus $T = T_1 + i T_2 = i R_1 R_2$, where $T_1=0$ and $T_2 = R_1 R_2$ is the volume of $T^2$. Plugging this form into the result and labeling the moduli/radii as $R_i$ immediately gives equation \ref{eq:squaretori}.

\end{document}